\documentclass[preprint,12pt]{elsarticle}

\usepackage{amssymb}

\begin{document}

\begin{frontmatter}

\title{Energy bands of atomic monolayers of various materials:
Possibility of energy gap engineering}

\author[sangi,kouku]{Tatsuo~Suzuki\corref{cor1}}
\ead{tatsuo@acp.metro-cit.ac.jp}

\author[kouku]{Yushi~Yokomizo}

\cortext[cor1]{Corresponding author}

\address[sangi]{Tokyo Metropolitan College of Industrial Technology, 8-17-1, Minami-Senju, Arakawa-ku, Tokyo, 116-0003, Japan}

\address[kouku]{Tokyo Metropolitan College of Aeronautical Engineering, 8-17-1, Minami-Senju, Arakawa-ku, Tokyo, 116-0003, Japan}

\begin{abstract}

The mobility of graphene is very high because the quantum Hall effects can be observed even
at room temperature.
Graphene has the potential of the material for novel devices because of this high mobility.
But the energy gap of graphene is zero, so graphene can not be applied to semiconductor devices
such as transistors, LEDs, etc.
In order to control the energy gaps, we propose atomic monolayers which consist of
various materials besides carbon atoms.

To examine the energy dispersions of atomic monolayers of various materials, we calculated
the electronic states of these atomic monolayers using density functional theory
with structural optimizations.
The quantum chemical calculation software {\it Gaussian 03} was used under periodic boundary conditions.
The calculation method is  LSDA/6-311G(d,p), B3LYP/6-31G(d), or B3LYP/6-311G(d,p).

The calculated materials are C (graphene), Si (silicene), Ge, SiC, GeC, GeSi,
BN, BP, BAs, AlP, AlAs, GaP, and GaAs.
These atomic monolayers can exist in the flat honeycomb shapes.
The energy gaps of these atomic monolayers take various values.
Ge is a semimetal;
AlP, AlAs, GaP, and GaAs are indirect semiconductors;
and others are direct semiconductors.

We also calculated the change of energy dispersions accompanied by the substitution of the atoms.
Our results suggest that the substitution of impurity atoms for monolayer materials can control
the energy gaps of the atomic monolayers.

We conclude that atomic monolayers of various materials have the potential for novel devices.

\end{abstract}

\end{frontmatter}


\section{Introduction}

Graphene \cite{2001} is an atomic monolayer which consists of carbon atoms arranged in a honeycomb lattice.
The mobility of graphene is very high because the quantum Hall effects can be observed even
at room temperature \cite{2002}.
Graphene has the potential of the material for novel devices because of this high mobility.
But the energy gap of graphene is zero \cite{2001}, so graphene can not be applied to semiconductor devices
such as transistors, LEDs, etc.
In order to control the energy gaps, we propose atomic monolayers which consist of
various materials besides carbon atoms.
Theoretical works show that the energy gaps of silicene which is the atomic monolayer
of silicon atoms \cite{2016,2043,2006}
and the atomic monolayer of germanium atoms \cite{2016} are zeros,
but a pioneering experimental work about atomic monolayers of boron nitride formed
on the metal surfaces \cite{2041ex} shows the finite energy gaps.
So we hope that other atomic monolayers also have the finite energy gaps
and have the potential for novel devices.
To examine the energy dispersions of atomic monolayers of various materials, we calculated
the electronic states of these atomic monolayers using density functional theory
with structural optimizations.
A honeycomb lattice was assumed as a crystal structure (Fig.1).
The atoms were alternately placed like Fig.1(b) in the case of a binary compound.

\section{Computational Method}
All calculations were performed using {\it Gaussian 03} package \cite{gaussian}.
{\it Gaussian} is a very famous quantum chemical calculation software, 
and has been used widely to calculate the electronic states of isolated molecules.
{\it Gaussian} was improved in 2003, and {\it Gaussian 03} has been able to calculate
the electronic states using periodic boundary conditions.
It can calculate the electronic states of not only three-dimensional crystals,
but also two-dimensional or one-dimensional materials.

We can specify the combination of method and basis set in {\it Gaussian 03} package.
We want to calculate the electronic states of the Kohn-Sham equations of
density functional theory, using the hybrid B3LYP exchange and correlation
functionals combined with 6-311G(d,p) basis set (B3LYP/6-311G(d,p)).
But unfortunately most calculations using B3LYP/6-311G(d,p) did not converge.
So we used local spin density approximation (LSDA) combined with 6-311G(d,p)
basis set (LSDA/6-311G(d,p)) or B3LYP combined with 6-31G(d) basis set (B3LYP/6-31G(d)) instead.
The adopted combinations of method and basis set are noted in Table 1.

We carried out the calculations of all electrons,
so the contributions from core electrons are included.

Furthermore the merits of our calculations are structural optimizations,
and the translation vectors are also changed after every self-consistent calculation.

\section{Results and discussion}

\subsection{A IV group element}

We show the energy dispersions of atomic monolayers which consist of a IV group element
in the periodic table in Fig.2.
We plot the lowest conduction bands and the highest valence bands
from other calculations by colored lines.

Fig.2(a) is a result of graphene which consists of carbon atoms.
The conduction band and the valence band of graphene contacts at K points,
which is famous as {\it Dirac points}.
The highest valence band of our calculation agrees very well
with other calculations \cite{2045ex,2033,2017,2046},
but the energy of the conduction band of our calculation at $\Gamma$ point
is rather higher than the results of other calculations.
We cannot understand this reason at present.
This is our problem to be solved.

Fig.2(b) is a result of silicene which consist of silicon atoms.
The conduction band and the valence band of silicene also contacts at K points as {\it Dirac points}.
Our result of silicene agrees very well with the solution of the local density approximation (LDA)
with the projected augumented wave (PAW) method \cite{2016}
and the solution of LDA in a plane wave basis set with Vanderbilt ultrasoft pseudopotentials \cite{2043}.
The bond length of our calculation is 2.23 \r{A}, the bond length of LDA
with the PAW method is 2.23 \r{A} \cite{2016}, and the bond length of LDA with Vanderbilt ultrasoft pseudopotentials is 2.22 \r{A} \cite{2043}.
These bond lengths also agree very well.
On the other hand, our result of silicene does not agree with the solutions of
tight-binding model \cite{2006}.

Fig.2(c) is a result of an atomic monolayer which consists of germanium atoms.
The conduction band and the valence band also contacts at K points,
but the bottom of conduction band at $\Gamma$ point is a little lower than the Fermi energy.
So the atomic monolayer of germanium is a semimetal.
This result agrees very well with the solution of LDA with the PAW method \cite{2016}.
The bond length of our calculation is 2.31 \r{A} and the bond length of
LDA with the PAW method is 2.32 \r{A} \cite{2016}.
These bond lengths also agree very well.

\subsection{Binary compounds located in IV group elements}

We show the energy dispersions of atomic monolayers which consist of binary compounds
located in IV group elements in the periodic table in Fig. 3.

The energy dispersions of SiC (Fig.3(a)) and GeC (Fig.3(b)) resemble each other,
and the conduction band and the valence band does not contact at K points.
The energy gap of SiC is 3.526 eV and the energy gap of GeC is 3.160 eV.
These values correspond to the wave lengths of ultraviolet rays.

But the conduction band and the valence band of GeSi (Fig.3(c)) almost contact
at K points.
This energy dispersion resembles the energy dispersion of silicene (Fig.2(b)).

At present we cannot explain the physical reason why the conduction band
and the valence band do not contact at K points for SiC and GeC
while both bands contact for GeSi.

\subsection{Binary compounds located between III and V group elements}

We show the energy dispersions of atomic monolayers which consist of binary compounds
located between III and V group elements in the periodic table in Fig. 4.
We calculated AlN and GaN also, but the calculations of AlN and GaN did not converge.

The energy gap of BN (Fig.4(a)) is very large (6.377 eV).
There are experimental data of the valence band of atomic monolayers
of BN formed on the metal surfaces \cite{2041ex}.
Our result agrees well with this experimental band structures.

The energy dispersions of BP (Fig.4(b)) and BAs (Fig.4(c)) resemble each other.
The energy gap of BP is 1.912 eV, which corresponds to the wave length of an orange light.
The energy gap of BAs is 1.594 eV, which corresponds to the wave length of an infrared ray.

The energy dispersions of AlP (Fig.4(d)), AlAs (Fig.4(e)), GaP (Fig.4(f)) and GaAs (Fig.4(g))
resemble one another.
The bottoms of conduction bands are located at $\Gamma$ point,
so these are indirect semiconductors.

We cannnot understand the physical factor which determines the transition type (direct or indirect)
at present.

\subsection{Possibility of energy gap engineering}

We show that the change of energy dispersions accompanied by the substitution of the atoms in Fig.5.
In the case of graphene (Fig.5(a)) and silicene (Fig.5(e)), the conduction bands and the valence bands make {\it Dirac points} at K points,
so the energy gaps are zeros.
But when the 12.5\% of carbon atoms are replaced with Si atoms (Fig.5(b)),
the energy gap becomes open.
This fact suggests that the substitution of impurity atoms for monolayer materials can control
the energy gaps of the atomic monolayers.
These atomic monolayers are direct semiconductors,
and the energy gap of SiC corresponds to the wave length of an ultraviolet ray.
So these atomic monolayers can be applied to novel optical devices.

We wanted to show the relation between the gap energy and the concentration of the substituted atoms.
But by the difficulty of convergence, the calculation method is not unified.
So we cannot estimate the gap-concentration relation quantitatively.

\subsection{Flatness}

In our all converged calculations, all atoms are located on the same plane which is formed
by the two translation vectors.
So the atomic monolayers of various materials can exist in the flat honeycomb shapes.

\section{Conclusion}

We calculated the electronic states of atomic monolayers which consist of various materials
using density functional theory with structural optimizations.
The calculated materials are C (graphene), Si (silicene), Ge, SiC, GeC, GeSi,
BN, BP, BAs, AlP, AlAs, GaP, and GaAs.
These atomic monolayers can exist in the flat honeycomb shapes.
The energy gaps of these atomic monolayers take various values.
Ge is a semimetal;
AlP, AlAs, GaP, and GaAs are indirect semiconductors;
and others are direct semiconductors.

We also calculated the change of energy dispersions accompanied by the substitution of the atoms.
Our results suggest that the substitution of impurity atoms for monolayer materials can control
the energy gaps of the atomic monolayers.
So we conclude that atomic monolayers of various materials have the potential for novel devices.

Our important problem is to establish synthetic methods of these atomic monolayers.

\begin{table}[p]
\caption{The results of calculations.}
\begin{center}
\hspace*{-2cm}
\begin{tabular}{|l|l|l|l|r|r|l|} \hline
Materials & The      & The  & Transition & Energy & Bond   & Method    \\
          & top      & bottom     &      & gap    & length & and       \\
          & of       & of         &    & (eV)   & (\r{A})  & basis set \\
          & valence  & conduction &      &        &        &           \\
          & band     & band       &      &        &        &           \\ \hline \hline
C (graphene)  & K    & K & Direct  &  0.036 & 1.41 & LSDA/6-311G(d,p)  \\ \hline
Si (silicene) & K    & K & Direct  &  0.064 & 2.23 & LSDA/6-311G(d,p)  \\ \hline
Ge   & K    & $\Gamma$ & Semimetal & -0.444 & 2.31 & LSDA/6-311G(d,p)  \\ \hline
SiC  & K    & K        & Direct    &  3.526 & 1.79 & B3LYP/6-31G(d)    \\ \hline
GeC  & K    & K        & Direct    &  3.160 & 1.83 & B3LYP/6-31G(d)    \\ \hline
GeSi & K    & K        & Direct    &  0.275 & 2.28 & B3LYP/6-31G(d)    \\ \hline
BN   & K    & K        & Direct    &  6.377 & 1.45 & B3LYP/6-311G(d,p) \\ \hline
BP   & K    & K        & Direct    &  1.912 & 1.87 & B3LYP/6-31G(d)    \\ \hline
BAs  & K    & K        & Direct    &  1.594 & 1.93 & B3LYP/6-31G(d)    \\ \hline
AlP  & K    & $\Gamma$ & Indirect  &  3.453 & 2.28 & B3LYP/6-31G(d)    \\ \hline
AlAs & K    & $\Gamma$ & Indirect  &  2.938 & 2.34 & B3LYP/6-31G(d)    \\ \hline
GaP  & K    & $\Gamma$ & Indirect  &  3.054 & 2.23 & B3LYP/6-31G(d)    \\ \hline
GaAs & K    & $\Gamma$ & Indirect  &  2.475 & 2.29 & B3LYP/6-31G(d)    \\ \hline
\end{tabular}
\end{center}
\end{table}

\begin{figure}[p]
\begin{minipage}{0.4\textwidth}
\centering
\includegraphics[width=4cm,clip]{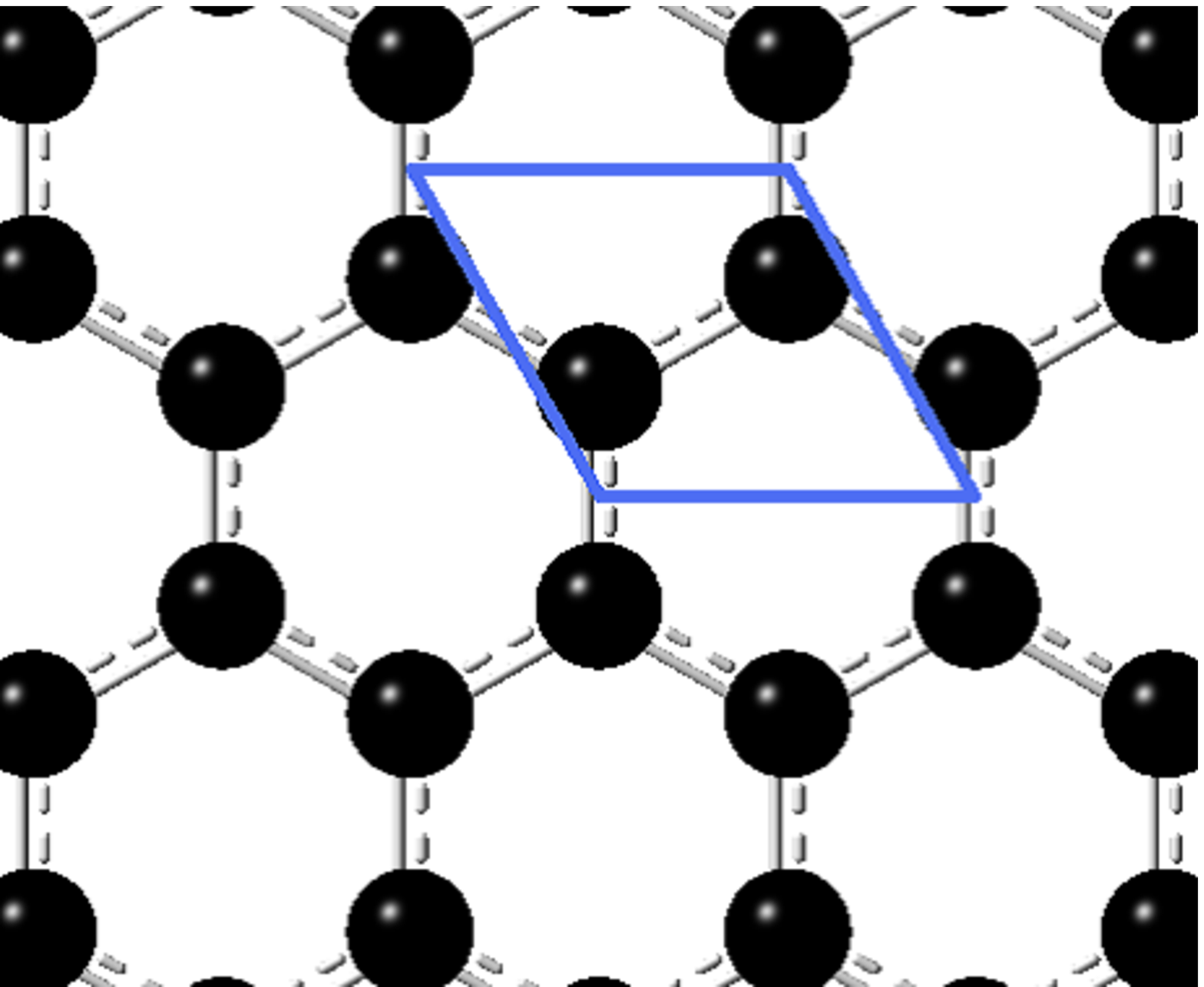} \\
(a) Element.
\end{minipage}
\begin{minipage}{0.4\textwidth}
\centering
\includegraphics[width=4cm,clip]{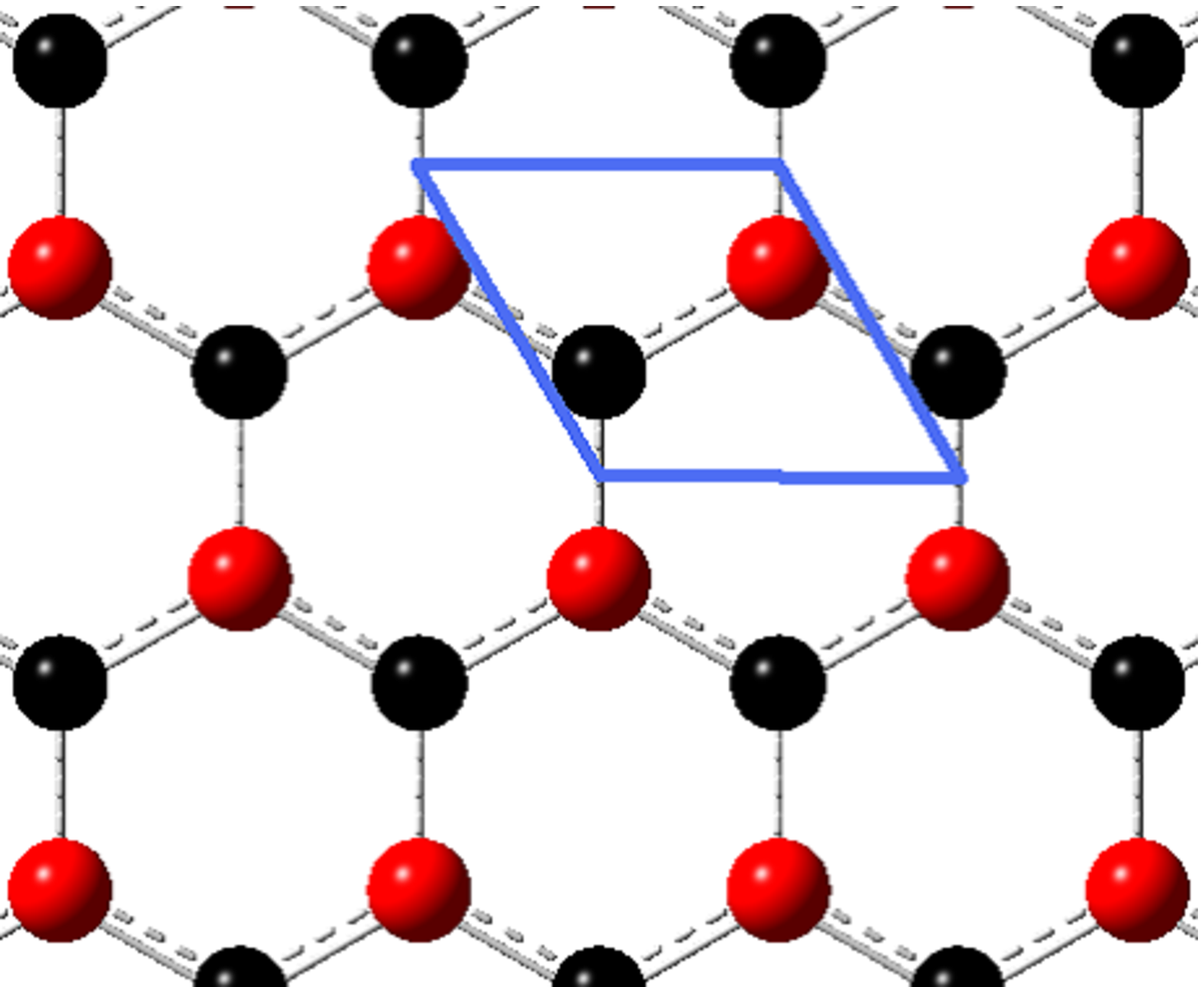} \\
(b) Binary compound.
\end{minipage}
\caption{A honeycomb lattice is assumed as a crystal lattice.
Blue lines indicate an unit cell.}
\end{figure}

\begin{figure}[p]
\includegraphics[width=5cm,clip]{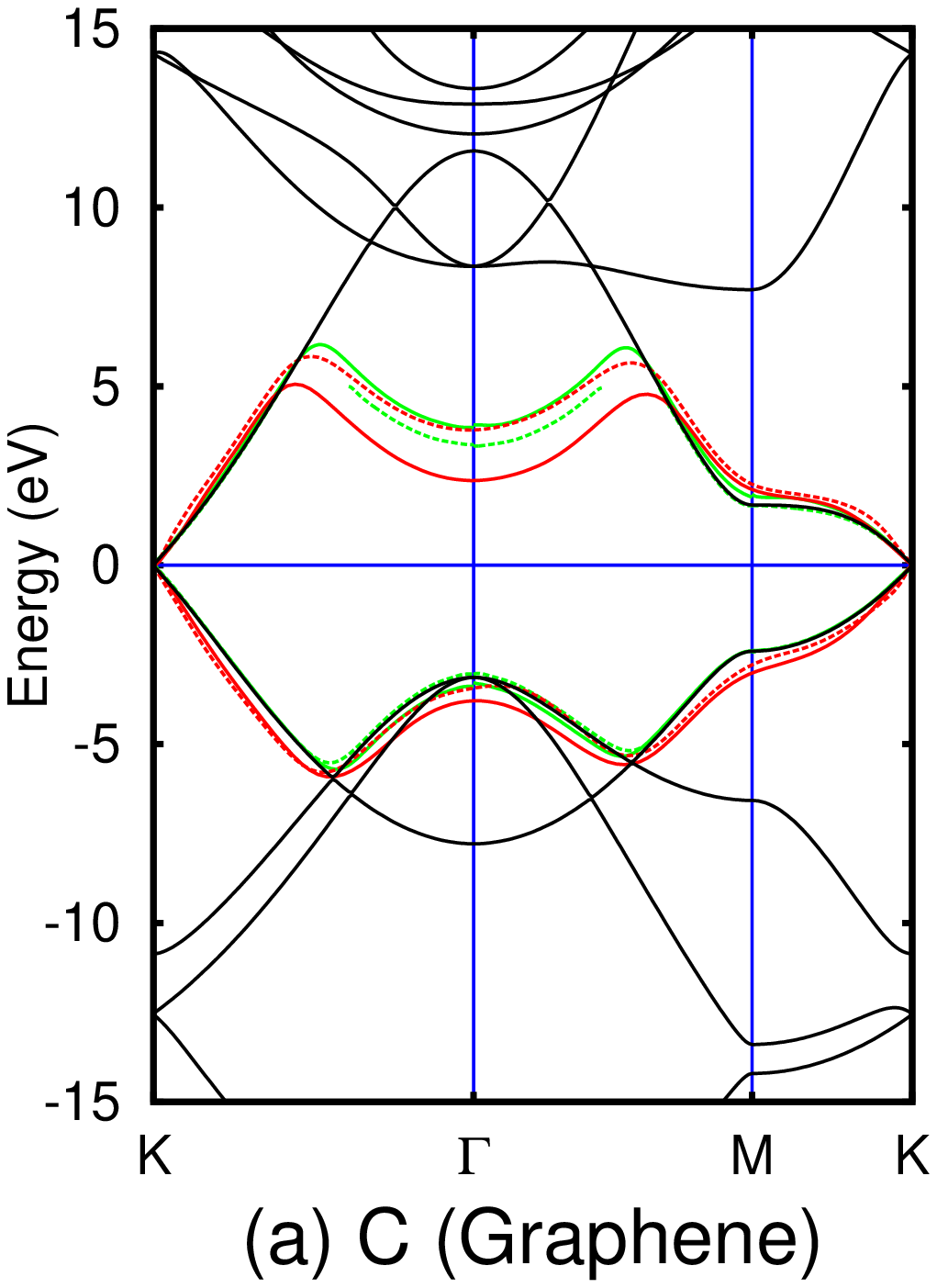} \hspace{-1cm}
\includegraphics[width=5cm,clip]{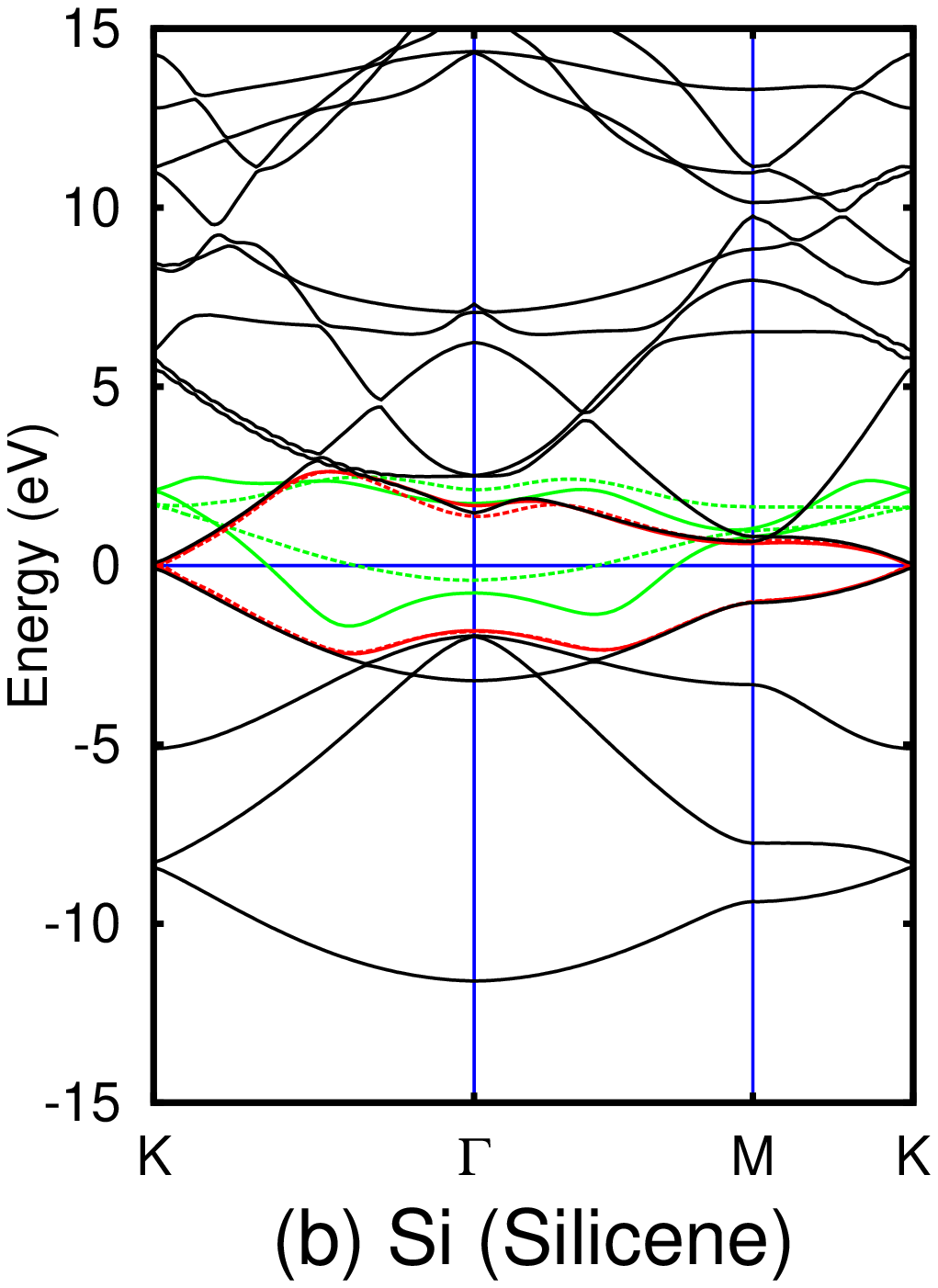} \hspace{-1cm}
\includegraphics[width=5cm,clip]{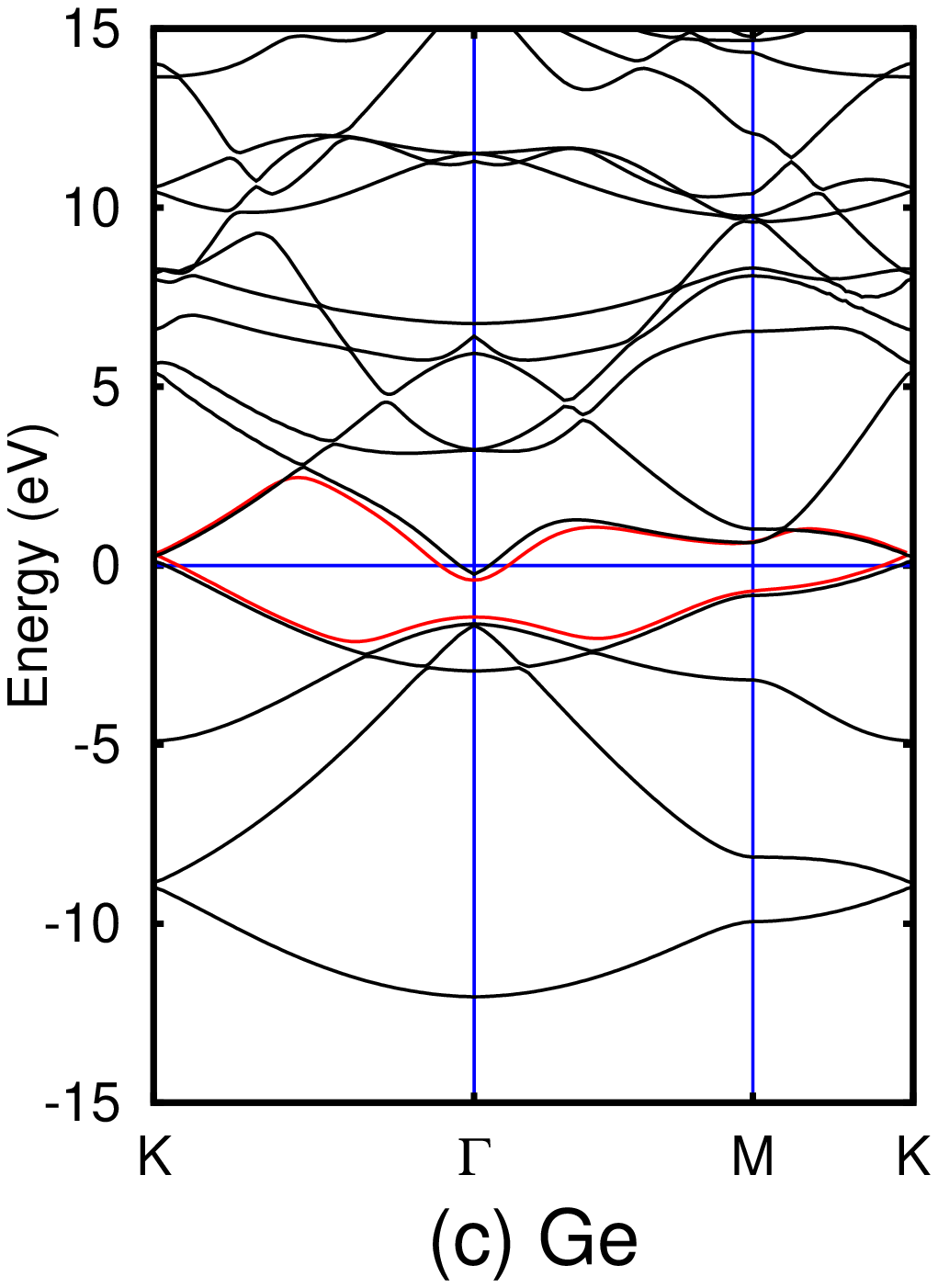}
\caption{Energy dispersions of atomic monolayers which consist of a IV group element in the periodic table.
The origin of energy is the Fermi energy.
Black solid lines are our calculations, whose calcutation method is  LSDA/6-311G(d,p).
In Fig. 2(a), both red solid lines \cite{2045ex} and red broken lines \cite{2033} are
the local density approximation (LDA) with the GW approximation,
green solid lines \cite{2017} are LDA with a full potential linear muffin orbital (FP-LMTO) method,
and green broken lines \cite{2046} are LDA with a norm-conserving pseudopotential.
In Fig. 2(b), red solid lines \cite{2016} are LDA with the projected augumented wave (PAW) method,
red broken lines \cite{2043} are LDA in a plane wave basis set with Vanderbilt ultrasoft pseudopotentials,
green solid lines \cite{2006} are a first-nearest neighbor (1NN) tight-binding model,
and green broken lines \cite{2006} are a second-nearest neighbor (2NN) tight-binding model.
In Fig. 2(c), red solid lines \cite{2016} are LDA with the PAW method.
}
\end{figure}

\begin{figure}[p]
\includegraphics[width=5cm,clip]{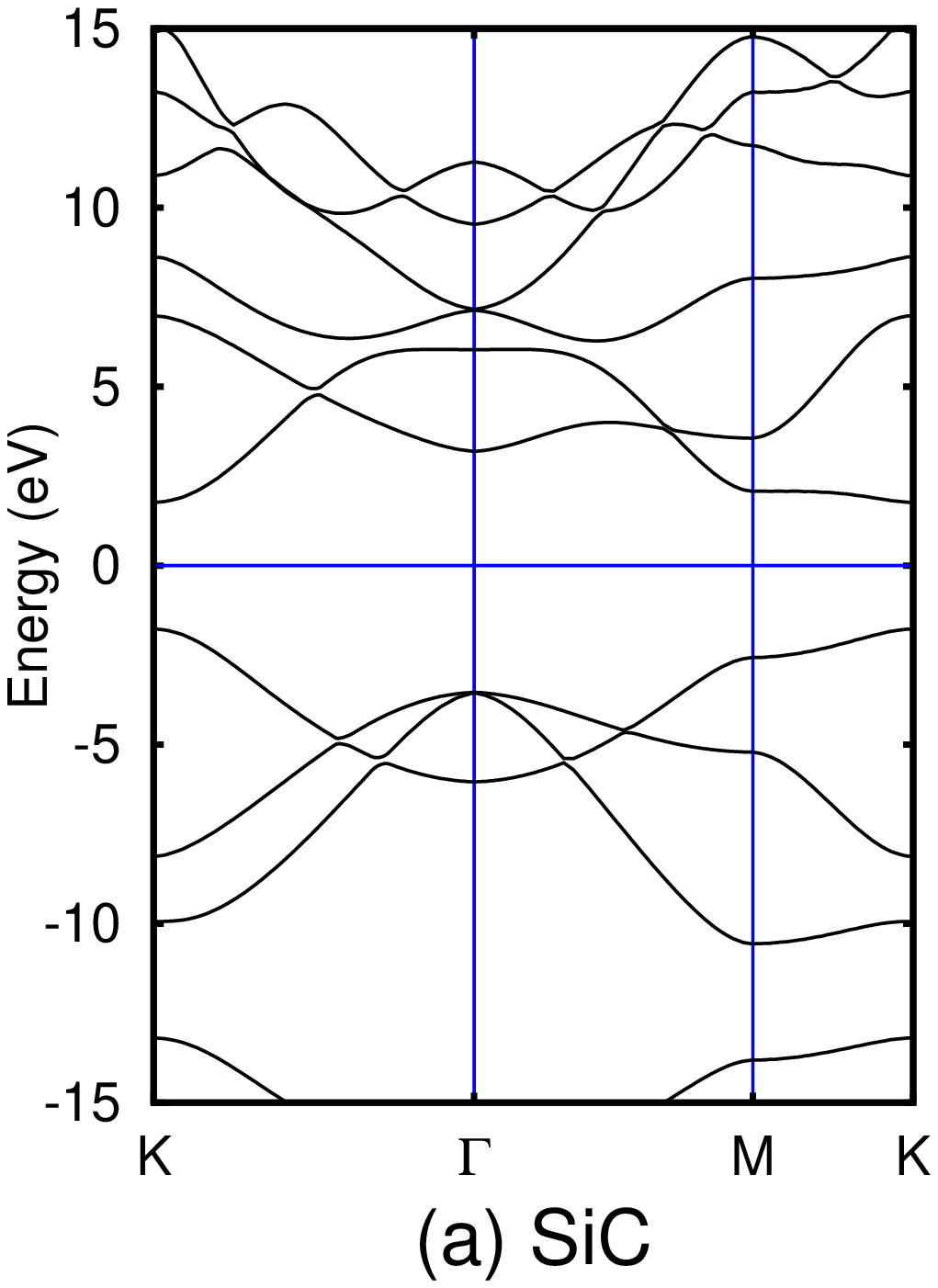} \hspace{-1cm}
\includegraphics[width=5cm,clip]{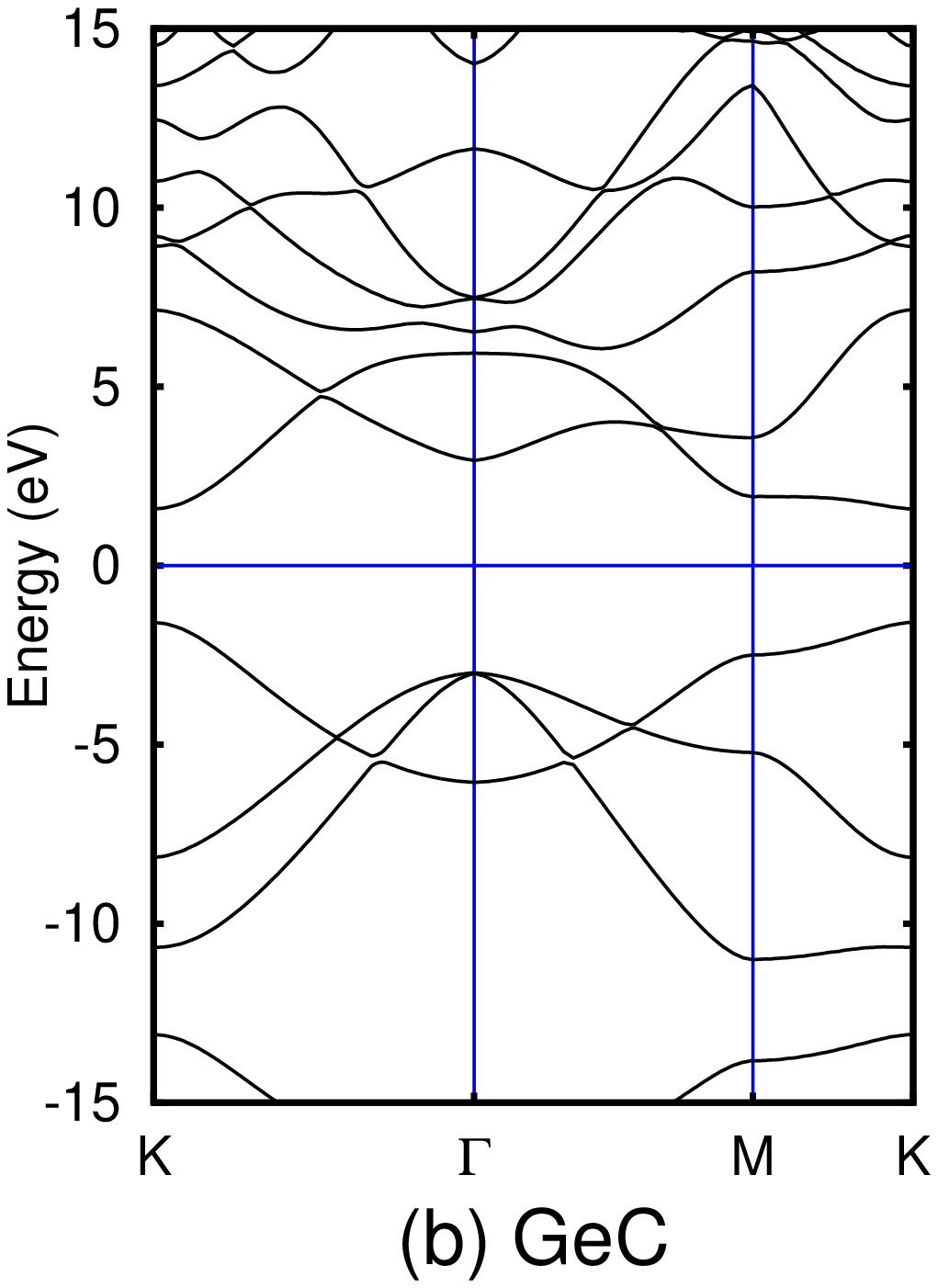} \hspace{-1cm}
\includegraphics[width=5cm,clip]{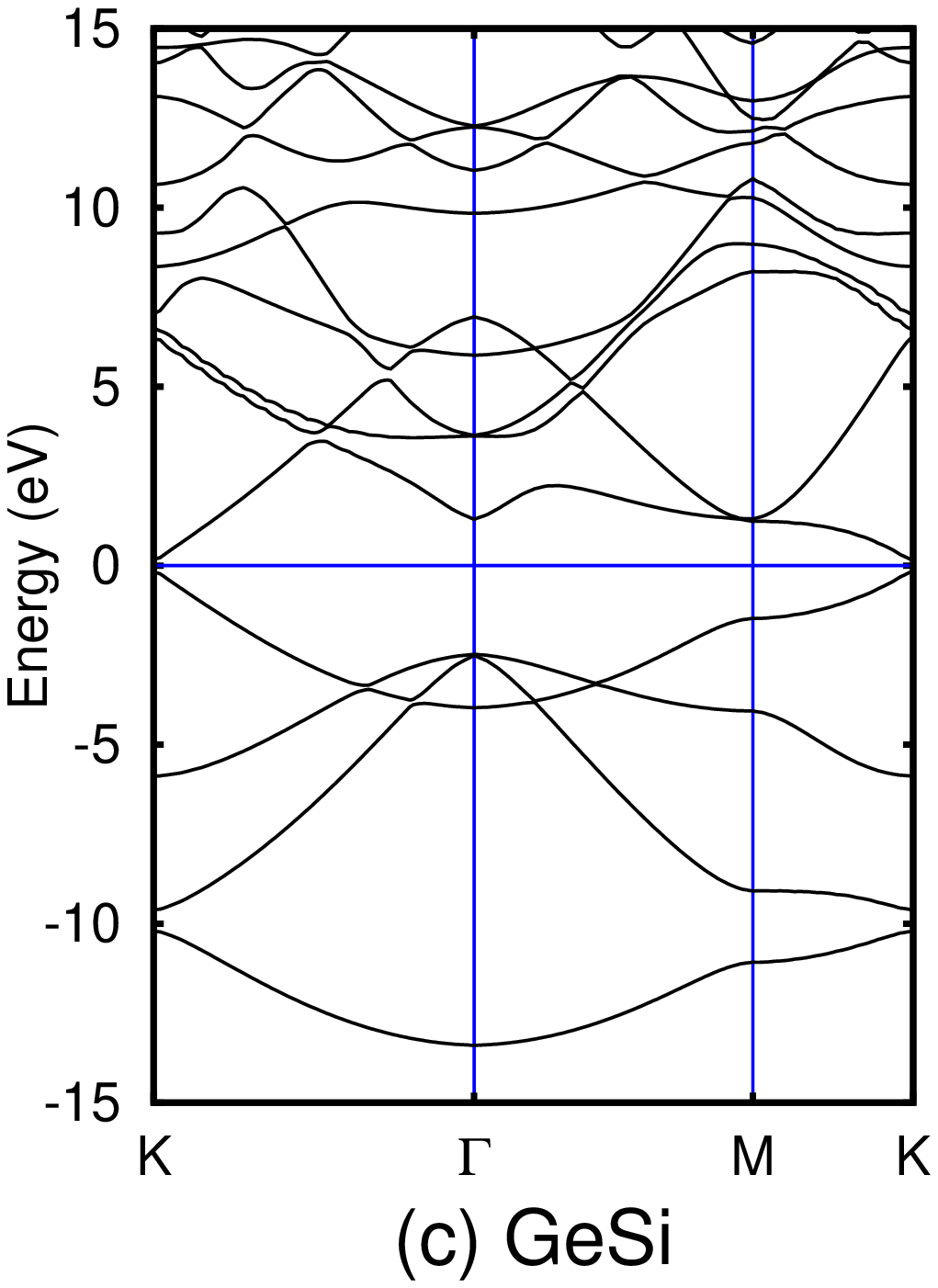}
\caption{Energy dispersions of atomic monolayers which consist of binary compounds located in IV group
elements in the periodic table.
The origin of energy is the Fermi energy.
The calcutation method is B3LYP/6-31G(d).}
\end{figure}

\begin{figure}[p]
\vspace*{-1cm}
\begin{minipage}{0.32\textwidth}
\includegraphics[width=5cm,clip]{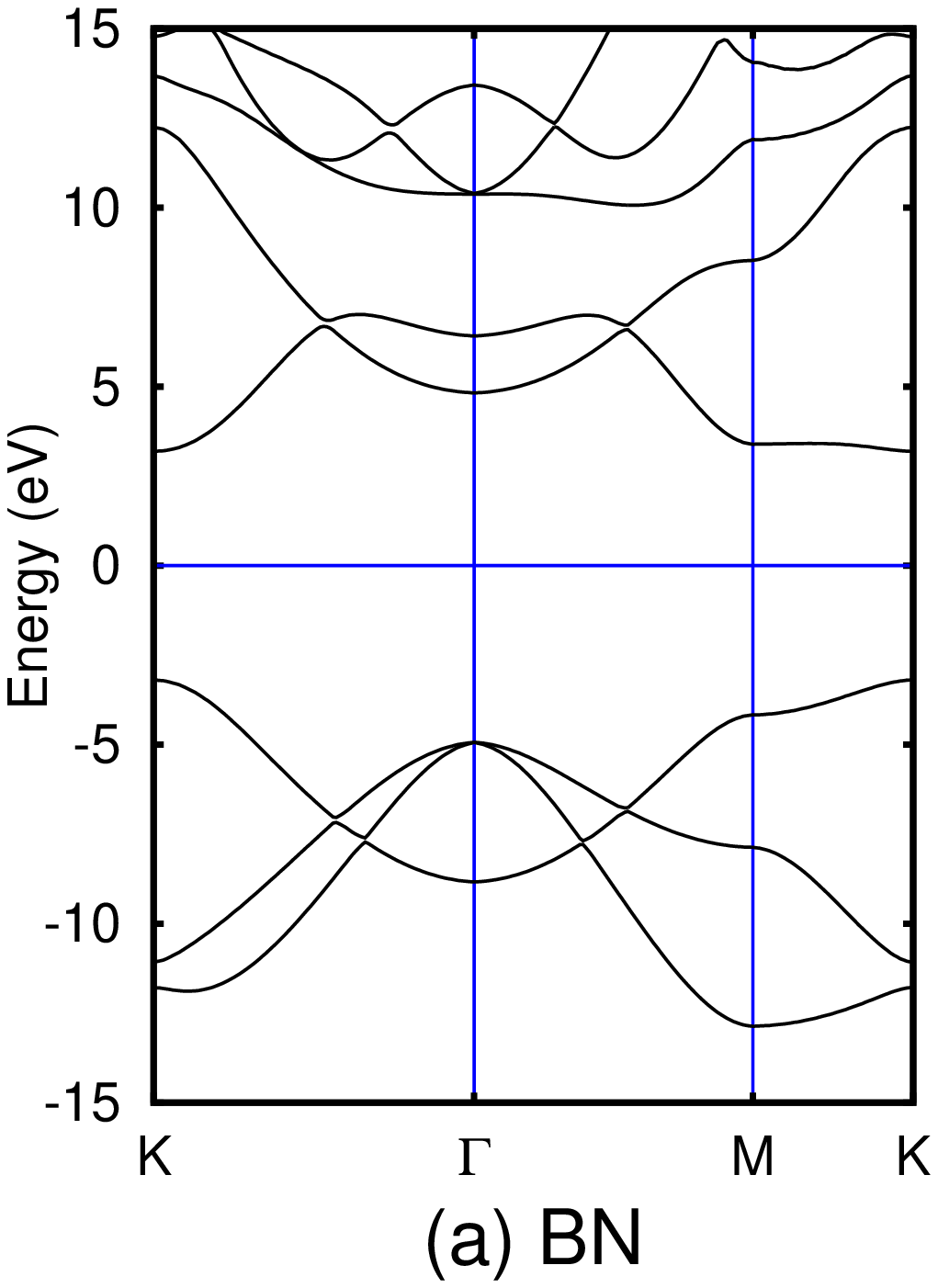}
\end{minipage}
\begin{minipage}{0.32\textwidth}
\includegraphics[width=5cm,clip]{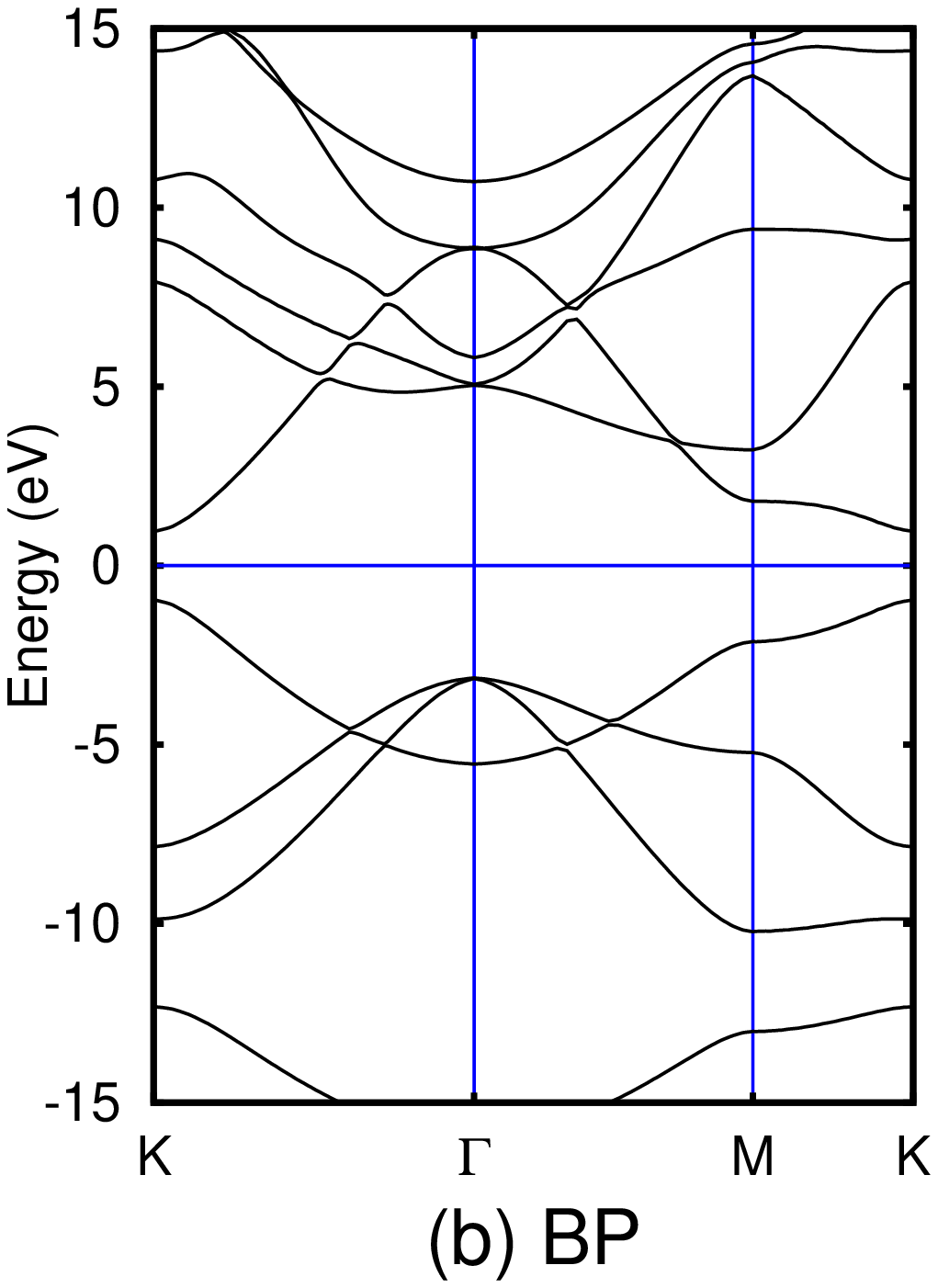}
\end{minipage}
\begin{minipage}{0.32\textwidth}
\includegraphics[width=5cm,clip]{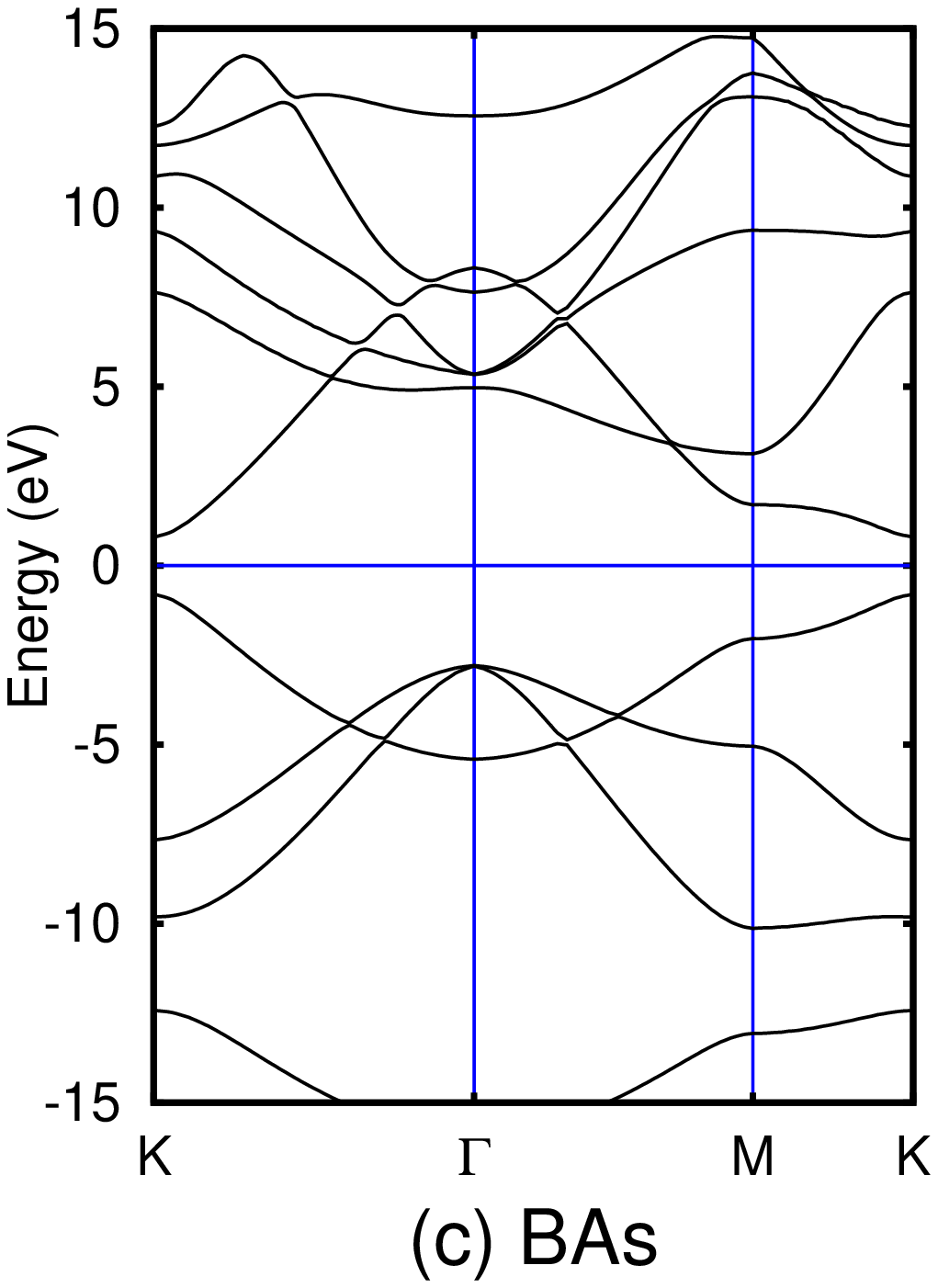}
\end{minipage} \\
\begin{minipage}{0.32\textwidth}
\hspace*{5cm}
\end{minipage}
\begin{minipage}{0.32\textwidth}
\includegraphics[width=5cm,clip]{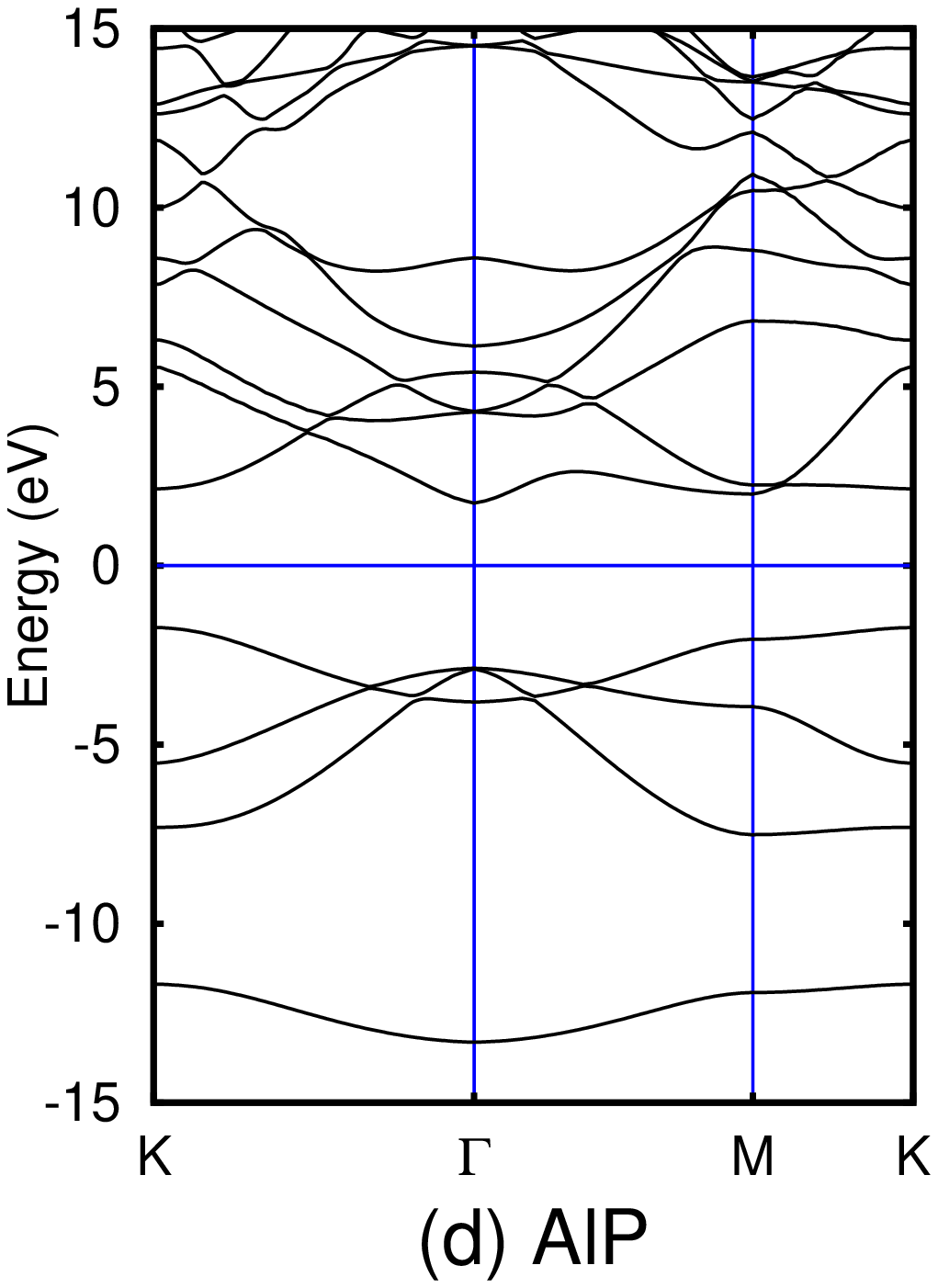}
\end{minipage}
\begin{minipage}{0.32\textwidth}
\includegraphics[width=5cm,clip]{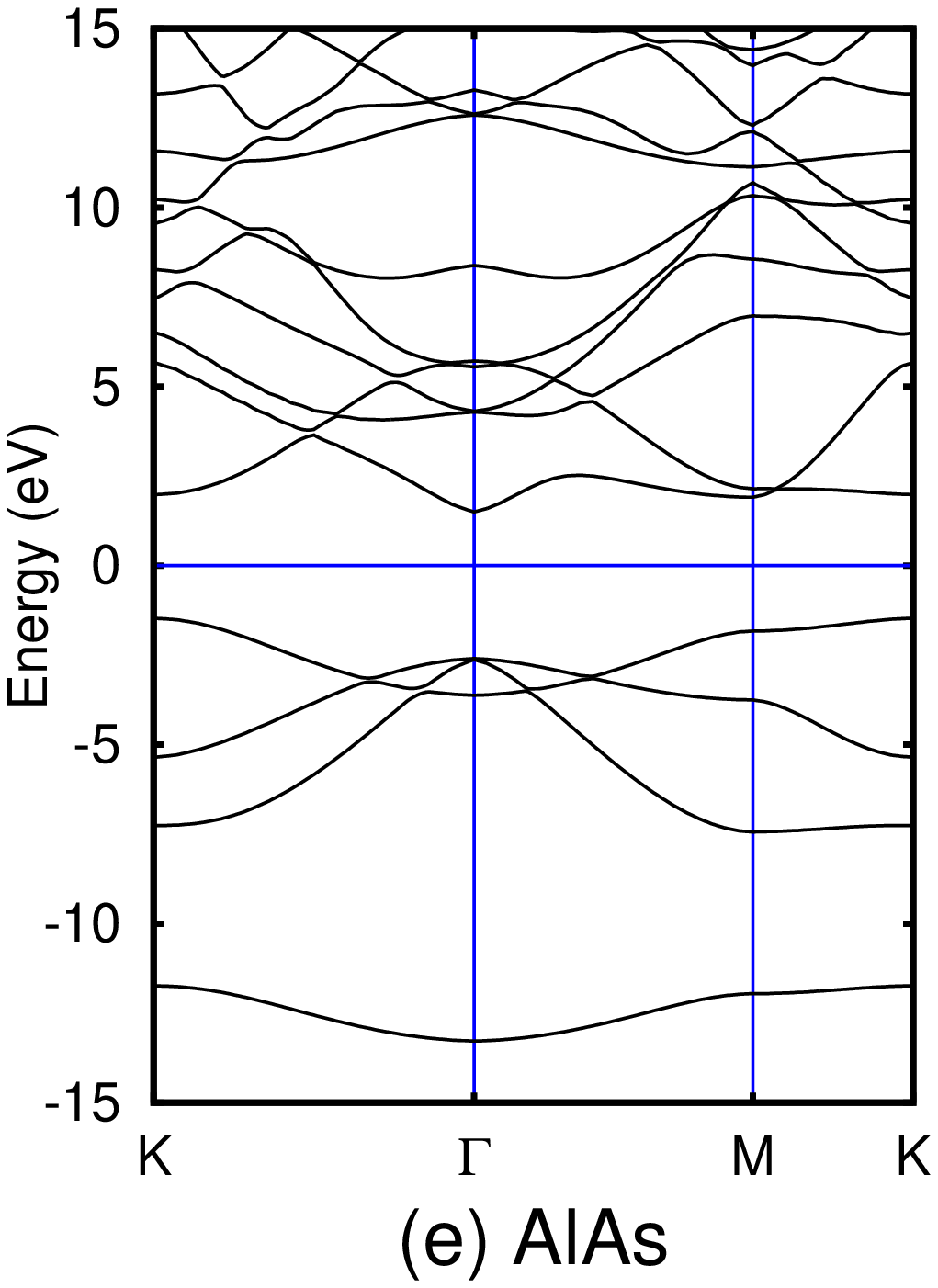}
\end{minipage} \\
\begin{minipage}{0.32\textwidth}
\hspace*{5cm}
\end{minipage}
\begin{minipage}{0.32\textwidth}
\includegraphics[width=5cm,clip]{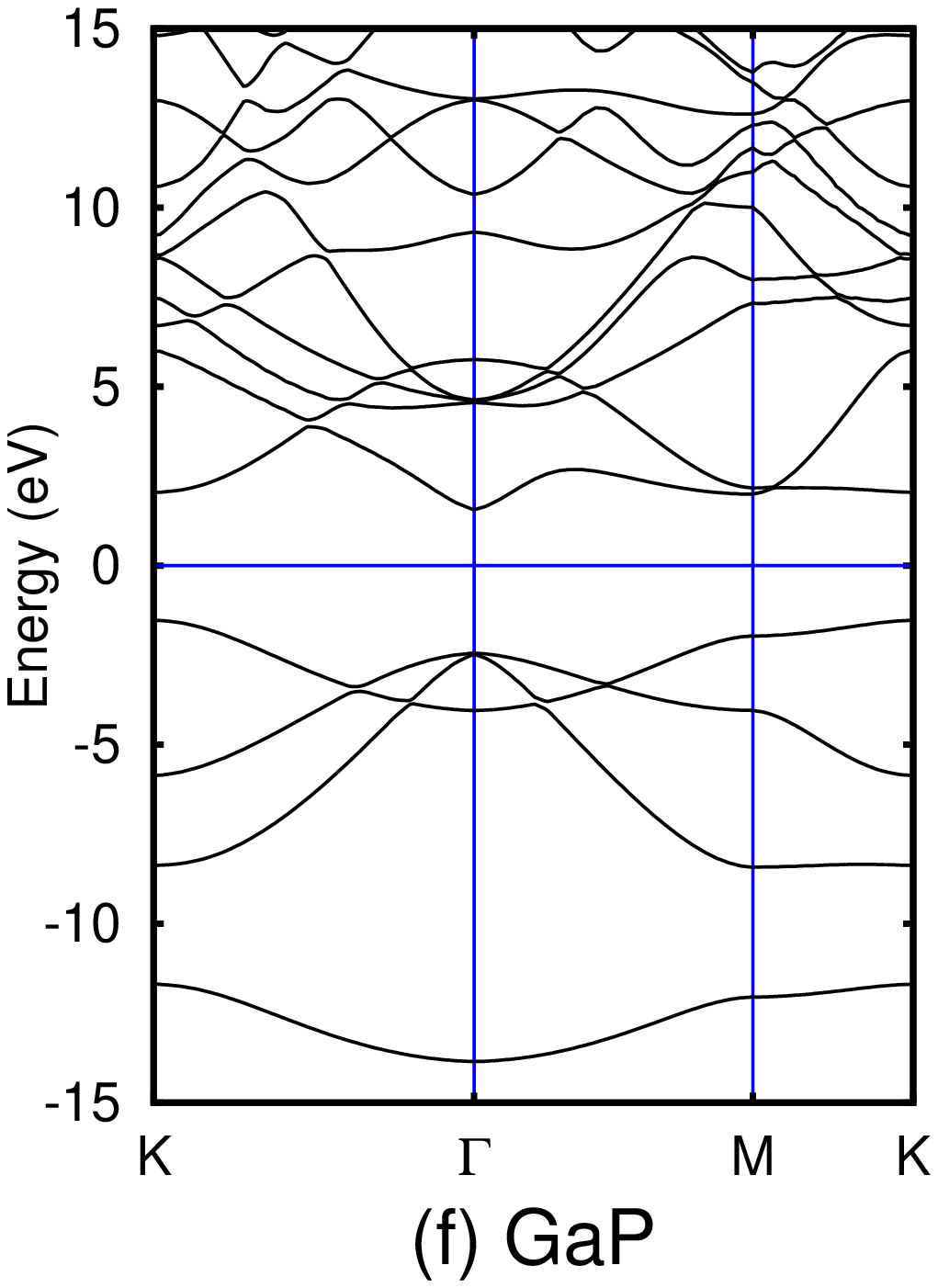}
\end{minipage}
\begin{minipage}{0.32\textwidth}
\includegraphics[width=5cm,clip]{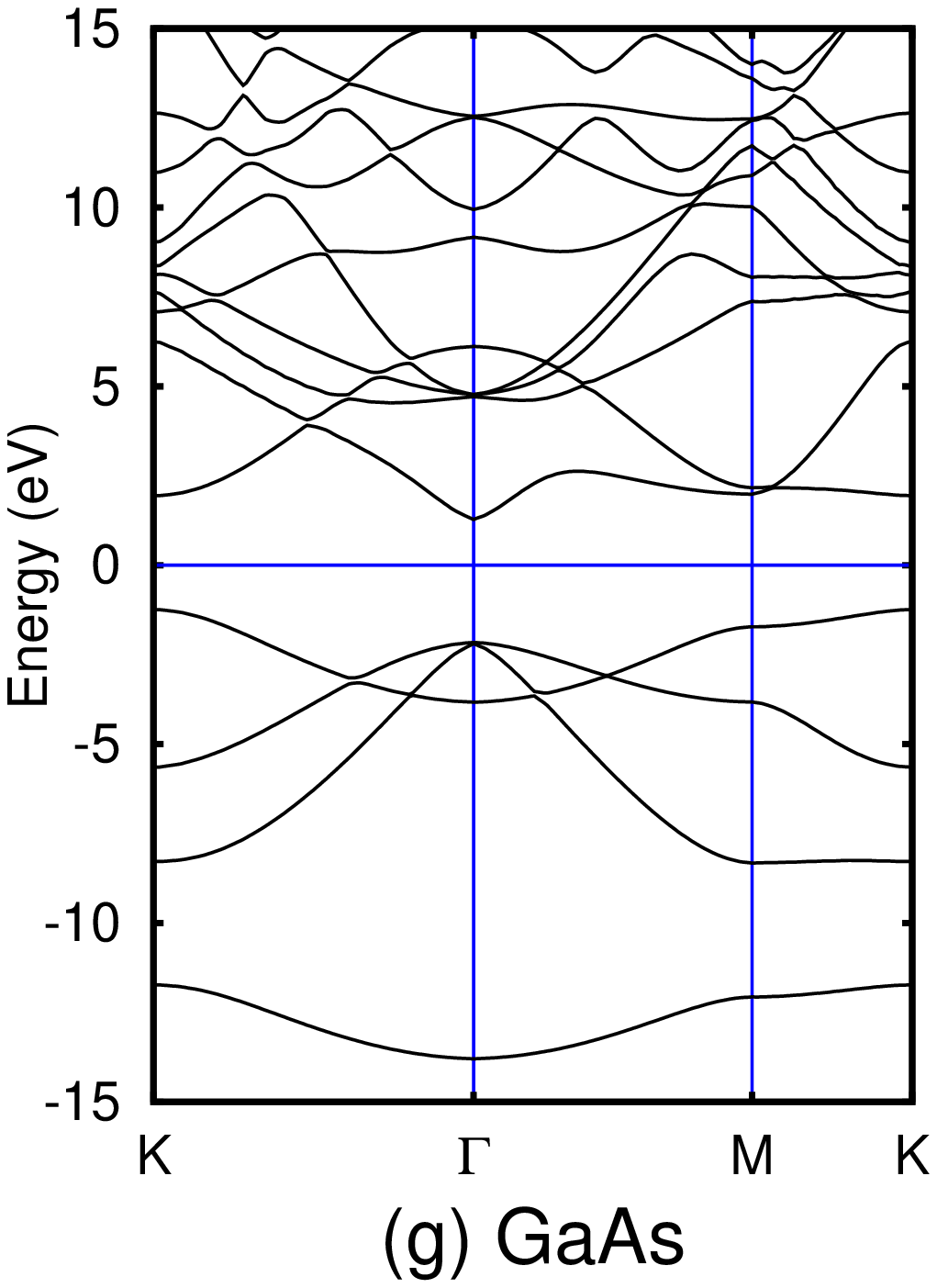}
\end{minipage}
\caption{Energy dispersions of atomic monolayers which consist of binary compounds located between III and V group elements in the periodic table.
The origin of energy is the Fermi energy.
The calcutation method of (a) is B3LYP/6-311G(d,p),
and the method of others is B3LYP/6-31G(d).}
\end{figure}

\begin{figure}[p]
\vspace*{-2.5cm}
\begin{minipage}{0.32\textwidth}
\hspace*{0.7cm}
\includegraphics[width=4cm,clip]{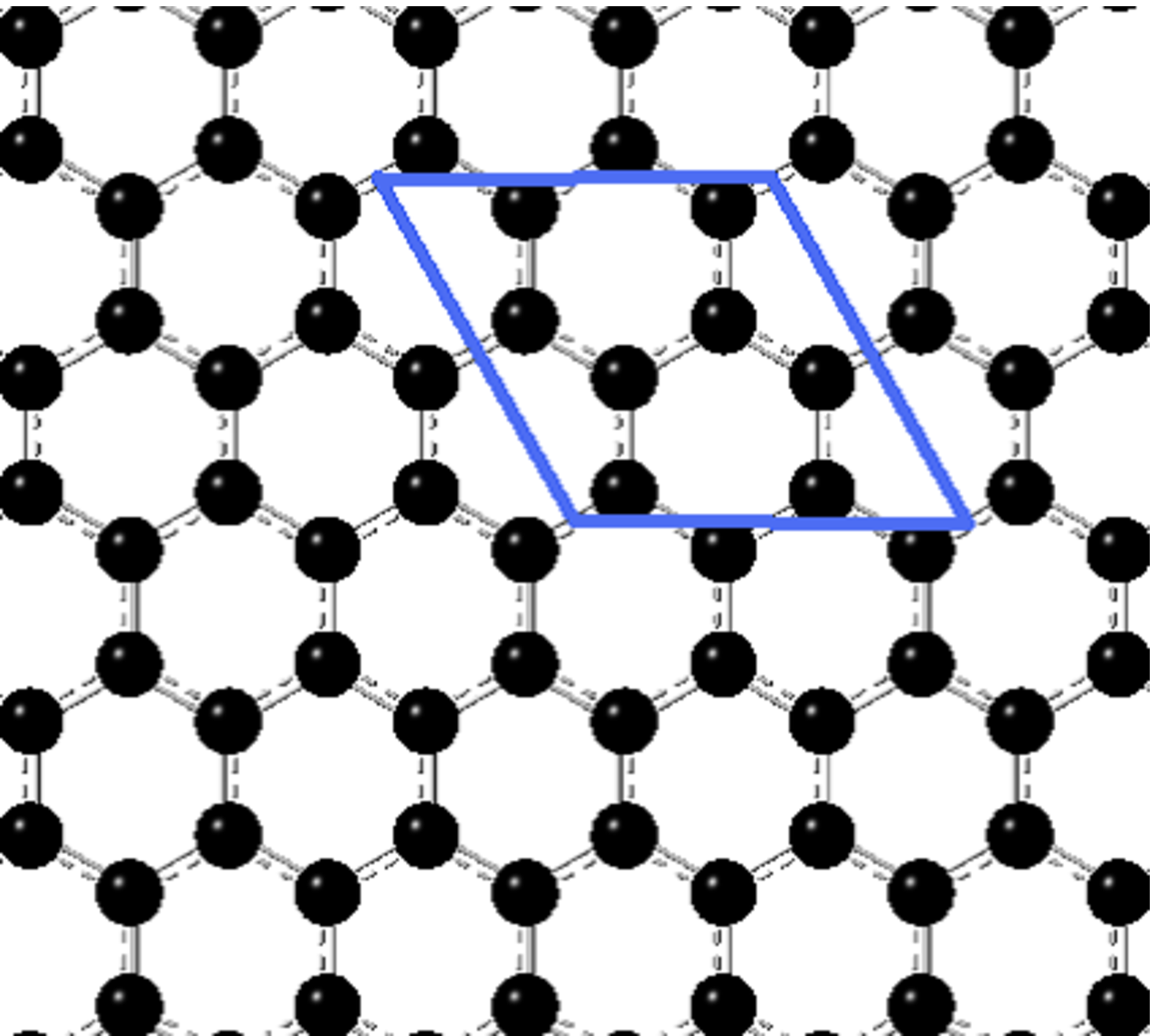}
\includegraphics[width=5cm,clip]{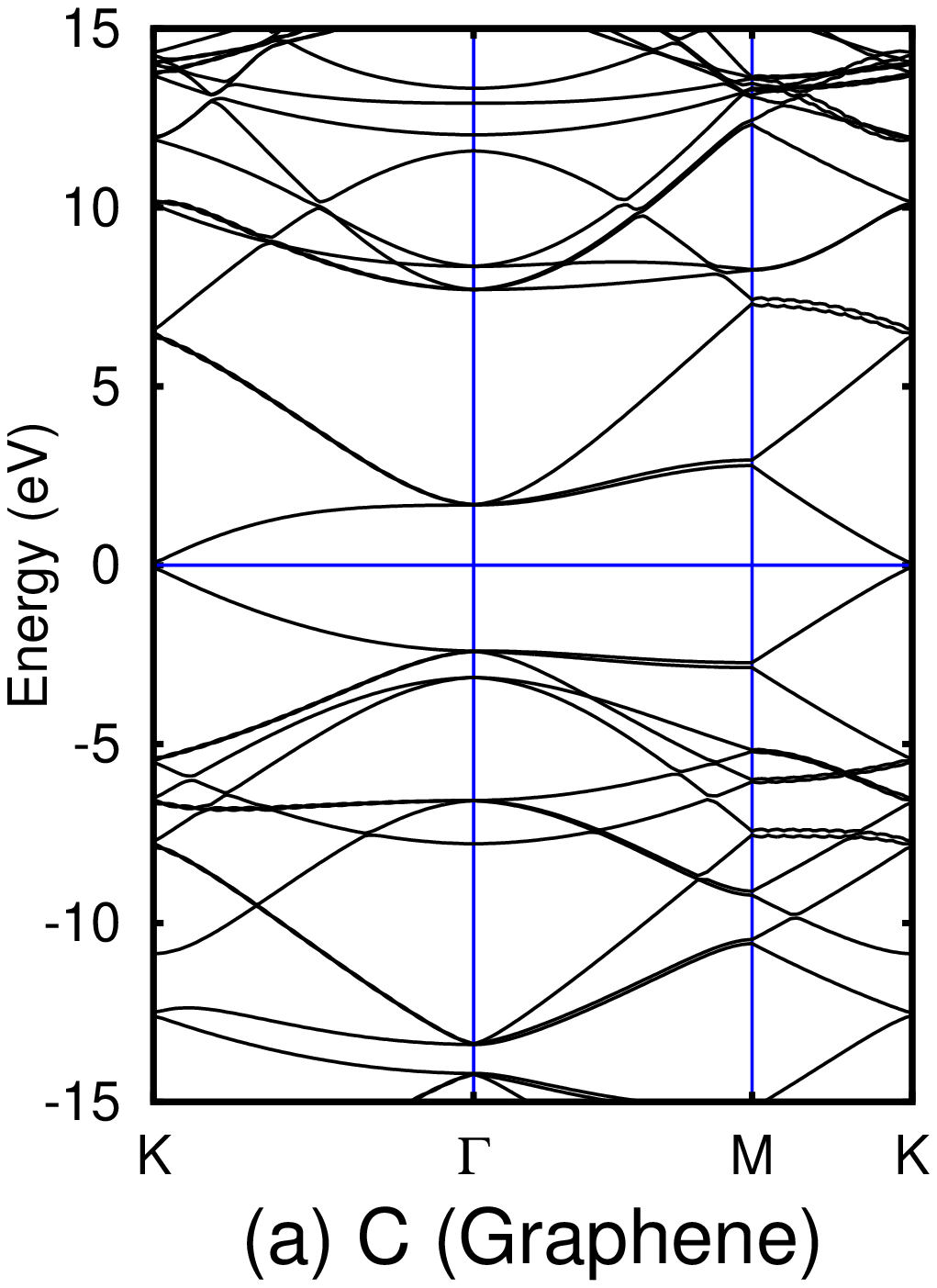}
\end{minipage}
\begin{minipage}{0.32\textwidth}
\hspace*{0.7cm}
\includegraphics[width=4cm,clip]{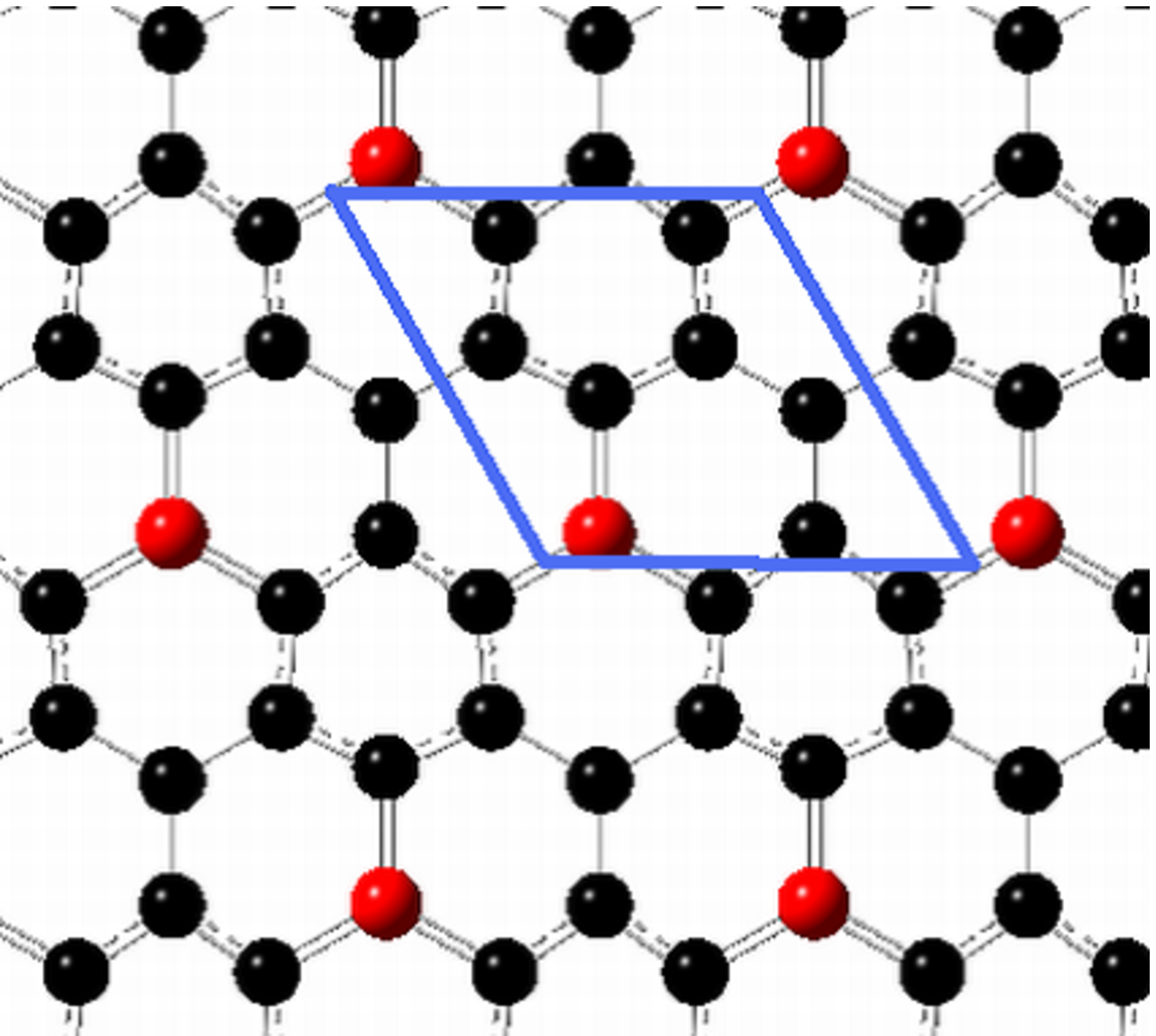}
\includegraphics[width=5cm,clip]{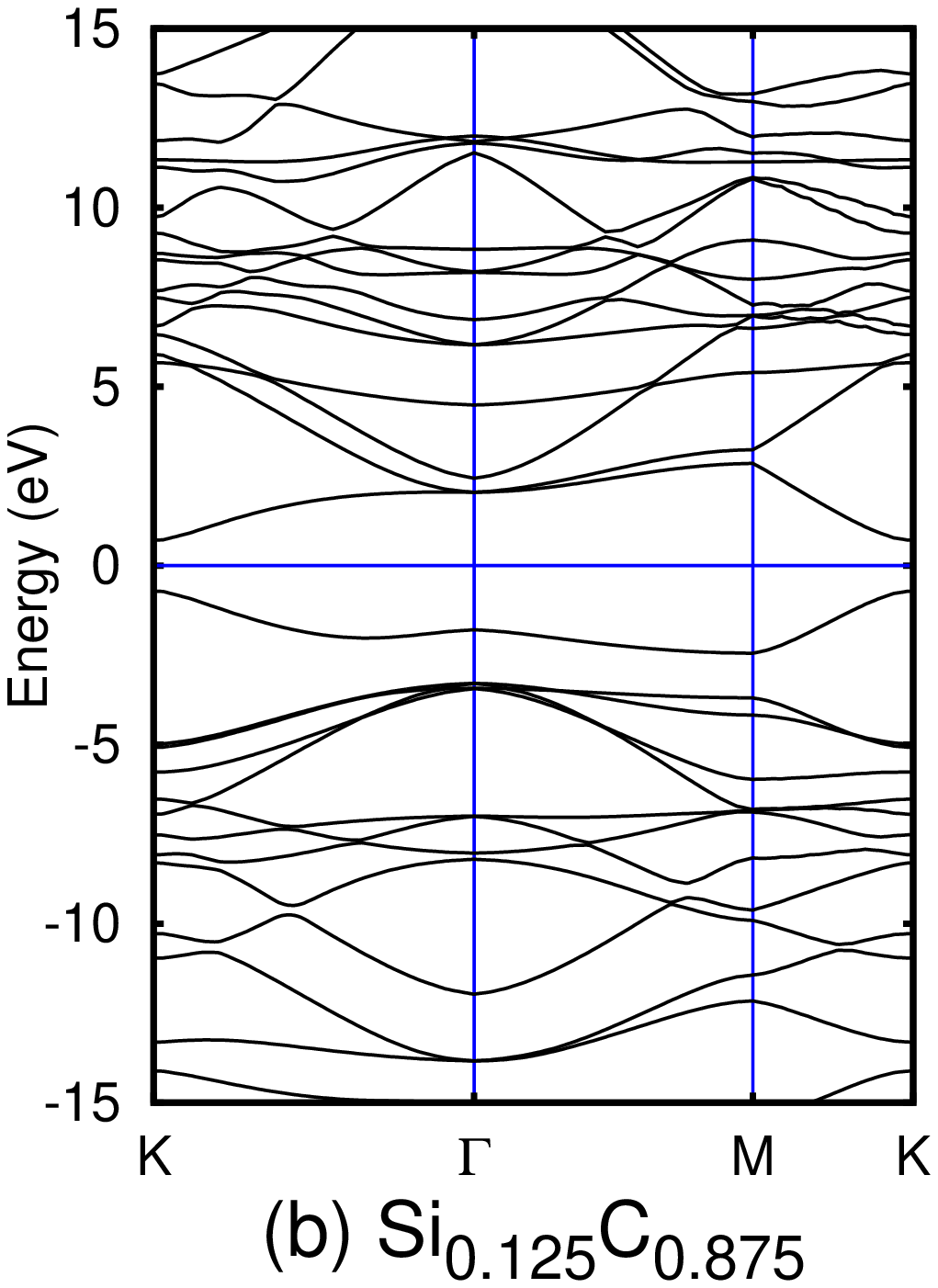}
\end{minipage}
\begin{minipage}{0.32\textwidth}
\hspace*{0.7cm}
\includegraphics[width=4cm,clip]{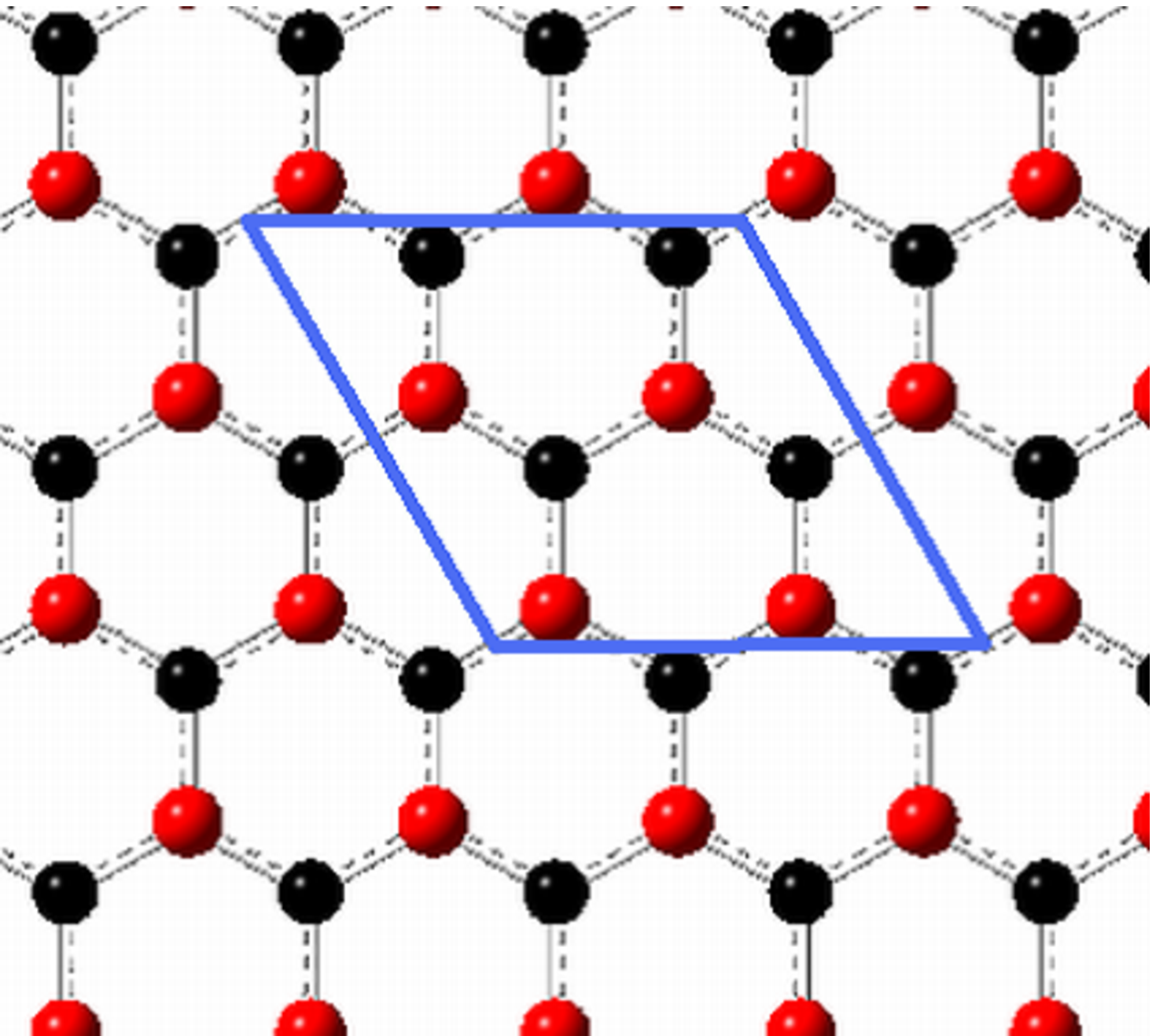}
\includegraphics[width=5cm,clip]{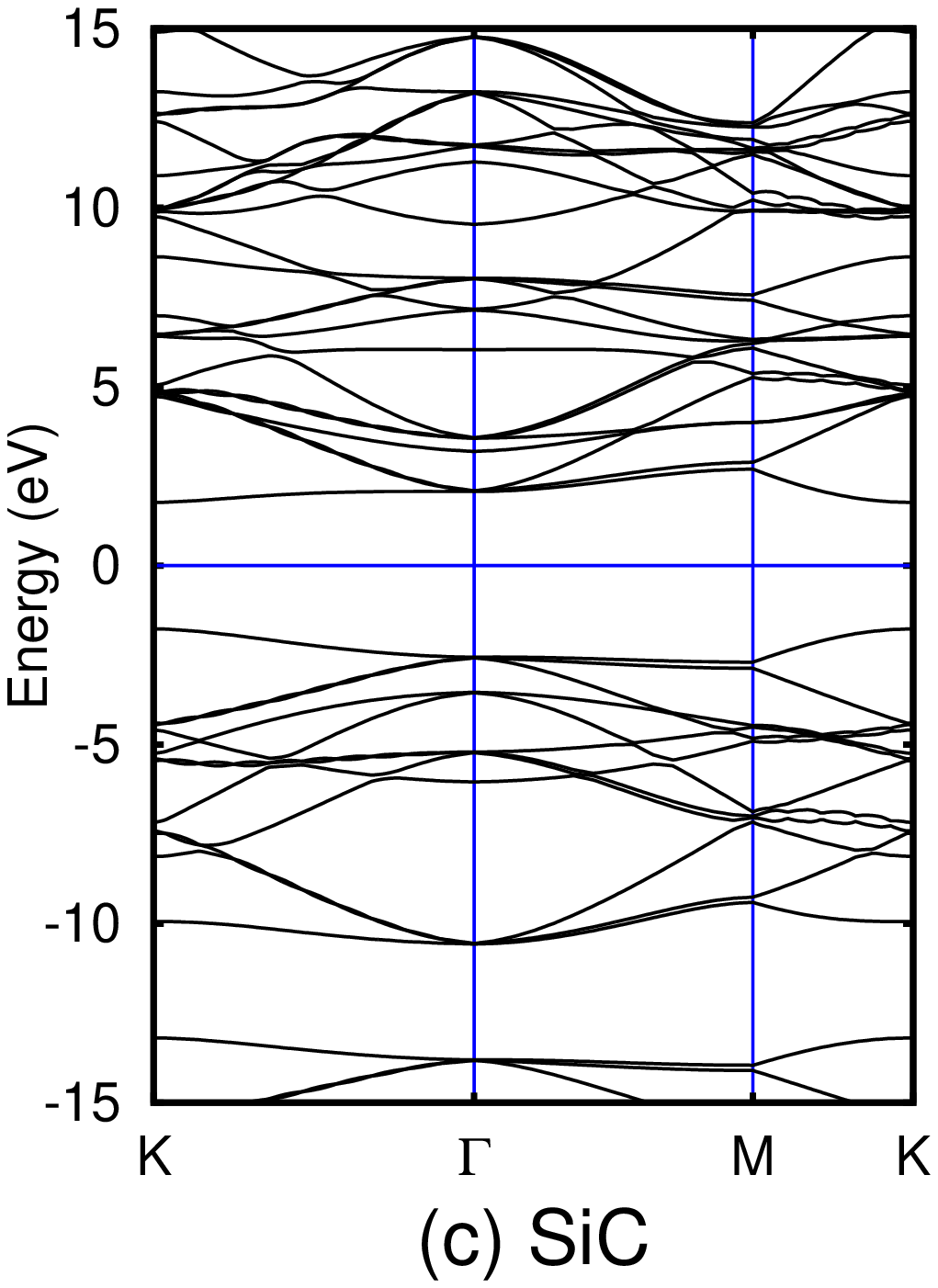}
\end{minipage} \\
\begin{minipage}{0.32\textwidth}
\hspace*{0.7cm}
\includegraphics[width=4cm,clip]{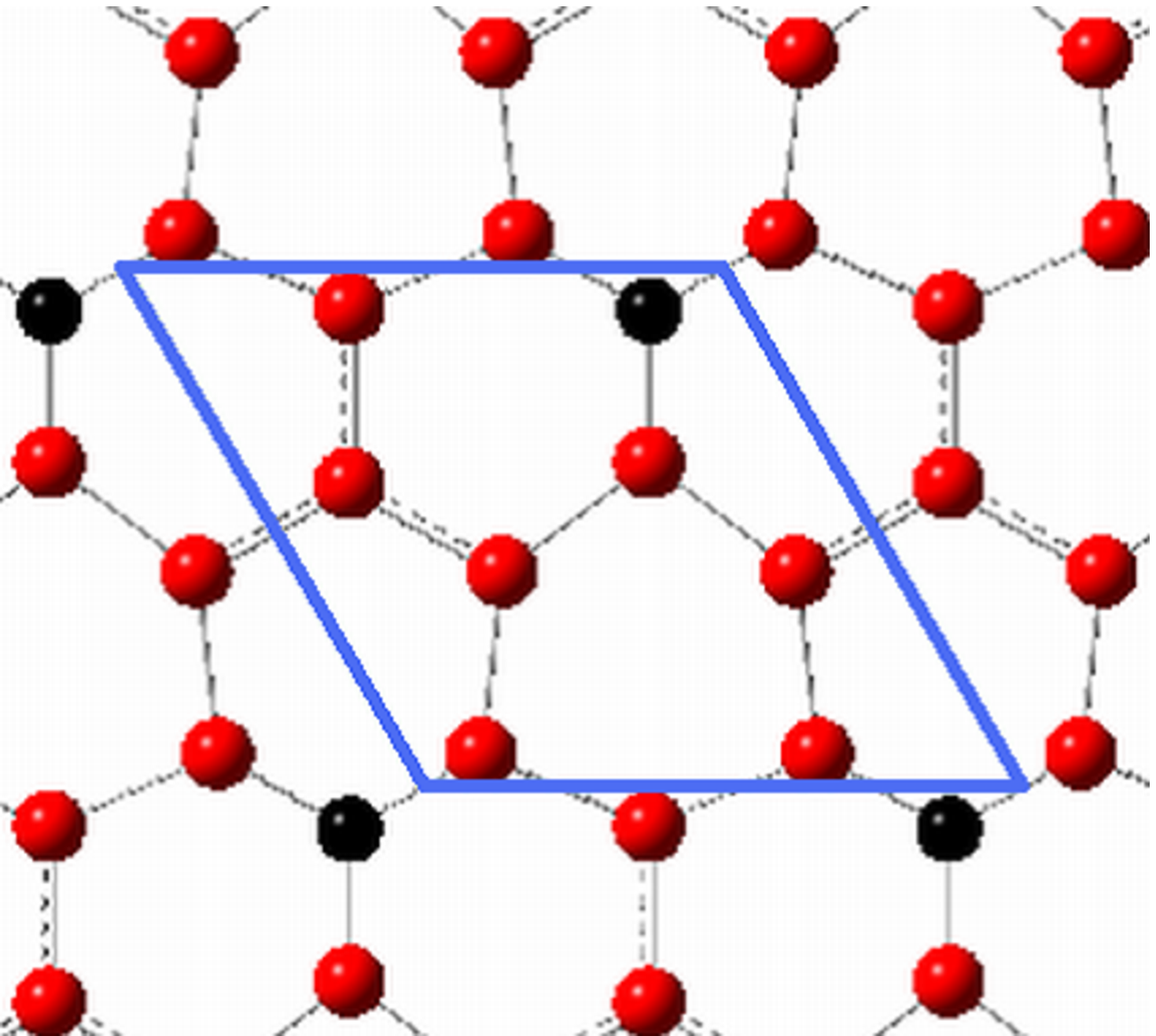}
\includegraphics[width=5cm,clip]{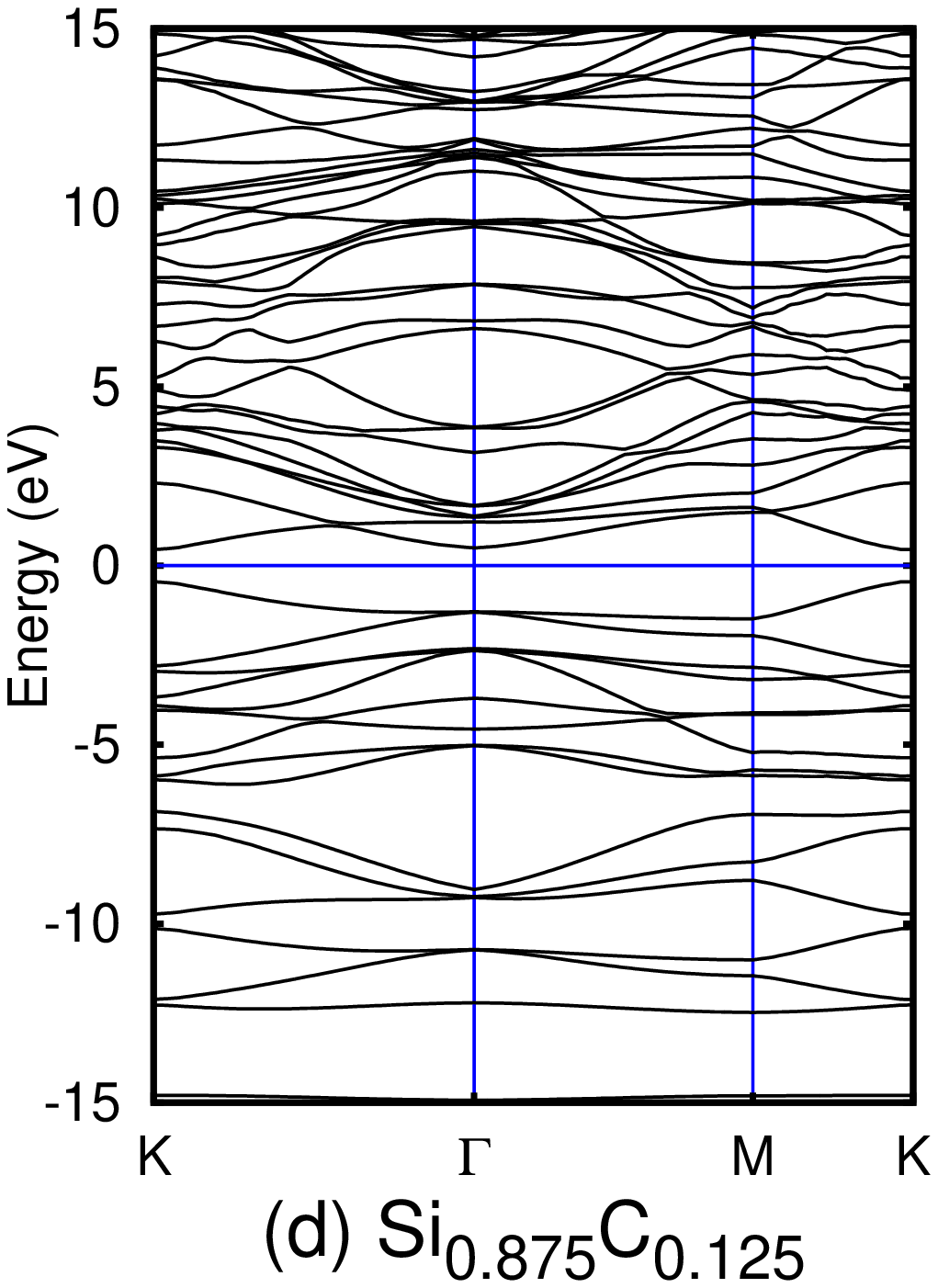}
\end{minipage}
\begin{minipage}{0.32\textwidth}
\hspace*{0.7cm}
\includegraphics[width=4cm,clip]{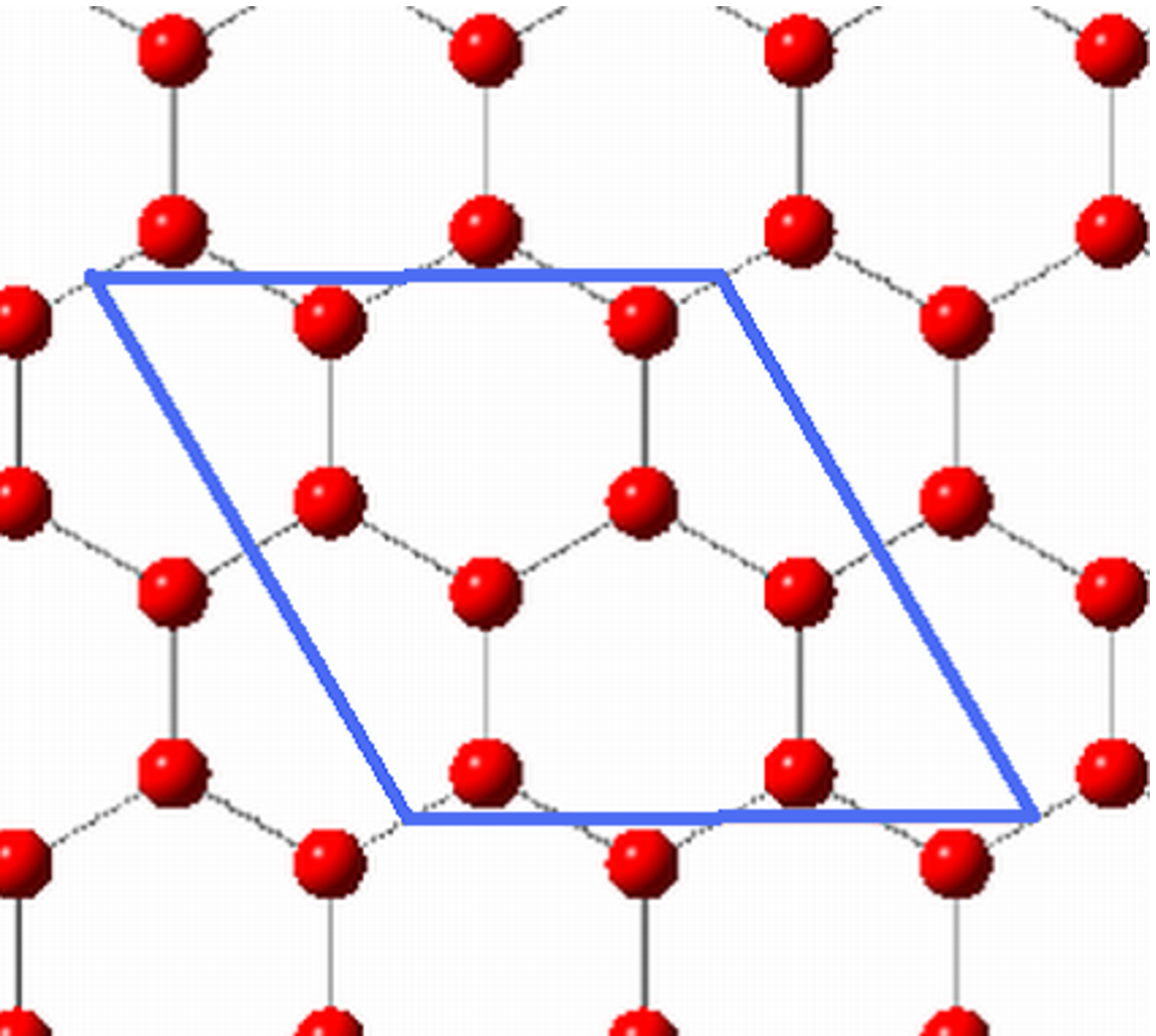}
\includegraphics[width=5cm,clip]{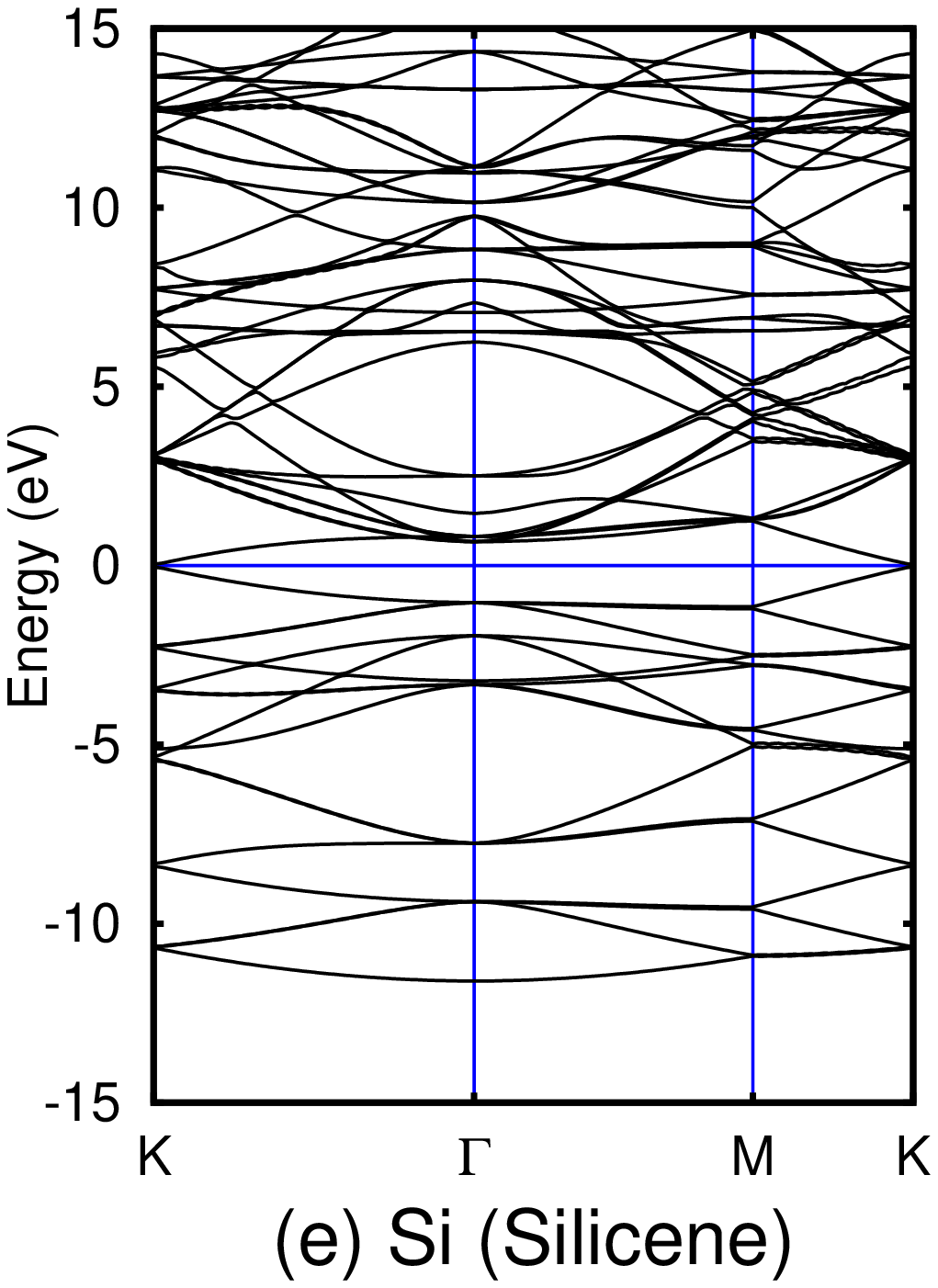}
\end{minipage}
\caption{{\it (Top)} The positions of atoms.
Black balls are carbon atoms, and red balls are silicon atoms.
Blue lines indicates an unit cell.
{\it (Bottom)} Energy dispersions of atomic monolayers which consist of carbon atoms and silicon atoms.
The origin of energy is the Fermi energy.
The calcutation method of (a) and (e) is LSDA/6-311G(d,p),
and the method of others is B3LYP/6-31G(d).}
\end{figure}

\end{document}